\RequirePackage{fix-cm}
\documentclass{svjour3}
\smartqed
\journalname{Empirical Software Engineering}

\usepackage{amsmath,amssymb,amsfonts}
\usepackage{algorithmic}
\usepackage{graphicx}
\usepackage{textcomp}
\usepackage{xcolor}
\usepackage{balance}
\usepackage{natbib}
\usepackage{listings}
\usepackage{xspace}
\usepackage{booktabs,tabularx,array}
\usepackage{morewrites}
\usepackage{footmisc}
\usepackage{rotating}
\usepackage{pdflscape}
\usepackage{rotating}
\graphicspath{.}

\usepackage{multibib}
\newcites{appA,appB,appC,appD,appE,appF,appG,appH,appI,appJ,appK,appL,appM,appN,appO,appP,appQ,appR}{%
2010 Selected Papers,%
2011 Selected Papers,%
2012 Selected Papers,%
2013 Selected Papers,%
2014 Selected Papers,%
2015 Selected Papers,%
2016 Selected Papers,%
2017 Selected Papers,%
2018 Selected Papers,%
2019 Selected Papers,%
2020 Selected Papers,%
2021 Selected Papers,%
2004 Selected Papers,%
2005 Selected Papers,%
2006 Selected Papers,%
2007 Selected Papers,%
2008 Selected Papers,%
2009 Selected Papers%
}

\usepackage[hidelinks,bookmarksdepth=-2]{hyperref}

\makeatletter
\let\old@lstKV@SwitchCases\lstKV@SwitchCases
\def\lstKV@SwitchCases#1#2#3{}
\makeatother
\usepackage{lstlinebgrd}
\makeatletter
\let\lstKV@SwitchCases\old@lstKV@SwitchCases

\lst@Key{numbers}{none}{%
    \def\lst@PlaceNumber{\lst@linebgrd}%
    \lstKV@SwitchCases{#1}%
    {none:\\%
     left:\def\lst@PlaceNumber{\llap{\normalfont
                \lst@numberstyle{\thelstnumber}\kern\lst@numbersep}\lst@linebgrd}\\%
     right:\def\lst@PlaceNumber{\rlap{\normalfont
                \kern\linewidth \kern\lst@numbersep
                \lst@numberstyle{\thelstnumber}}\lst@linebgrd}%
    }{\PackageError{Listings}{Numbers #1 unknown}\@ehc}}
\makeatother

\usepackage{colortbl}
\usepackage{makecell}
\usepackage[inline]{enumitem}

\newcommand{\figref}[1]{Figure~\ref{#1}\xspace}
\newcommand{\tabref}[1]{Table~\ref{#1}\xspace}
\newcommand{\secref}[1]{Section~\ref{#1}\xspace}
\newcommand{\appref}[1]{Appendix~\ref{#1}\xspace}

\definecolor{lightgray}{gray}{0.98}
\definecolor{codeblue}{rgb}{0.13,0.13,1}
\definecolor{codegrey}{rgb}{0.36,0.35,0.38}
\definecolor{codegreen}{rgb}{0,0.5,0}
\definecolor{codered}{rgb}{0.9,0,0}
\definecolor{OliveGreen}{cmyk}{0.64,0,0.95,0.40}
\definecolor{DarkGreen}{cmyk}{0.40,0,0.50,0.15}
\definecolor{purple}{cmyk}{0.41,0.73,0,0}
\definecolor{LightRed}{cmyk}{0,0.682,0.728,0}
\definecolor{LightLightBlue}{rgb}{0.87,0.92,0.968}
\definecolor{LightLightRed}{rgb}{0.98,0.9,0.84}
\definecolor{LightLightGreen}{rgb}{0.886,0.941,0.851}
\colorlet{codehl}{yellow!10}

\makeatletter
\lst@Key{countblanklines}{true}[t]%
    {\lstKV@SetIf{#1}\lst@ifcountblanklines}

\lst@AddToHook{OnEmptyLine}{%
    \lst@ifnumberblanklines\else%
       \lst@ifcountblanklines\else%
         \advance\c@lstnumber-\@ne\relax%
       \fi%
    \fi}
\makeatother

\lstdefinestyle{default}{
    escapechar=!,
	backgroundcolor=\color{lightgray},
	keywordstyle=\color{codeblue},
 	stringstyle=\color{codegreen},
	commentstyle=\color{codegrey},
	basicstyle=\footnotesize\ttfamily,
	columns=fullflexible,
    breaklines=true,
	frame=none,
    framexleftmargin=0.5em,
    framexrightmargin=0pt, 
	numbers=left,
	numberblanklines=false,
    countblanklines=false,
	stepnumber=1,
	numbersep=0.75em,
	numberstyle=\scriptsize,
	keepspaces=true,
    showspaces=false,
    showstringspaces=false,
	showtabs=false,
    tabsize=2,
}

\lstdefinestyle{boa}{
	mathescape=false,
	escapechar=\%,
	emphstyle=\color{purple},
	emphstyle={[2]\color{LightRed}},
	morestring=[b]`,
	gobble=0,
	morekeywords={_,argument,input,view,exists,foreach,ifall,output,of,weight,stop,visit,before,after,switch,default,current,view},
	emph={int,string,bool,table,float,time,array,stack,map,visitor,%
true,false,%
top,sum,mean,maximum,minimum,set,collection,bottom,%
Project,Person,CodeRepository,Revision,ChangedFile,ASTRoot,Namespace,Declaration,Type,Method,Variable,Statement,Expression,Modifier,IfStatement,%
BooleanLiteral,StringLiteral,PDG,%
InfixExpr,PrefixExpr,PostfixExpr,ForStatement,DoStatement,WhileStatement,TryStatement,ReturnStatement,ThrowStatement,ContinueStatement,BreakStatement,LabeledStatement,%
ExpressionKind,NEW,LITERAL,EQ,NEQ,OP_ADD,OP_MULT,OP_SUB,OP_INC,OP_DEC,BIT_NOT,LOGICAL_NOT,%
TypeKind,CLASS,ANONYMOUS,%
ModifierKind,OTHER,%
ChangeKind,DELETED,%
StatementKind,IF,BREAK,RETURN,THROW,CONTINUE,LABEL,FOR,DO,WHILE,SWITCH,TRY,%
RepositoryKind,SVN},
	emph={[2]logDead,isboollit,isstringlit,isinfix,isprefix,ispostfix,isfixingrevision,getast,iskind,hasfiletype,isliteral,getsnapshot,has_modifier_public,%
new,clear,values,contains,keys,add,format,def,len,match,lowercase,yearof,haskey,remove,strfind,push,pop,peek,normalize,getpdg,getcrypthash,gettotalnodes,gettotaledges,gettotalcontrolnodes,getpdgslice,_row},
}

\lstnewenvironment{Boa}
  {\lstset{language=bash,style=default,style=boa}}
  {}

\makeatletter
\newif\if@blinded

\@blindedfalse  

\if@blinded
  
\else
  
\fi
\makeatother

\usepackage{mdframed}
\newcounter{FindingCounter}
\stepcounter{FindingCounter}
\newcommand{\findings}[1]{
	\begin{mdframed}[backgroundcolor=gray!10,
			linewidth=0.75pt,
			roundcorner=5pt,
			innertopmargin=4pt,
			innerbottommargin=4pt,
			innerrightmargin=4pt,
			innerleftmargin=4pt,
			leftmargin=4pt,
			rightmargin=4pt,
			skipabove=2mm,
			skipbelow=0mm,
			font=\small]
		{\bf{Finding~\arabic{FindingCounter}:}}~ #1
	\end{mdframed}\vspace{2mm}
	\stepcounter{FindingCounter}
}

\begin{document}

\title{Pitfalls and Guidelines for Using Time-Based Git Data\thanks{This paper is a revised and extended version of \citet{msrstudy}.}}

\author{Samuel W. Flint \and Jigyasa Chauhan \and Robert Dyer}
\authorrunning{Flint, Chauhan, and Dyer}

\institute{Corresponding author: Samuel W. Flint \at
              University of Nebraska--Lincoln \\
              \email{swflint@huskers.unl.edu}
           \and
           Jigyasa Chauhan \at
              University of Nebraska--Lincoln\\
              \email{jchauhan2@huskers.unl.edu}
           \and
           Robert Dyer \at
              University of Nebraska--Lincoln\\
              \email{rdyer@unl.edu}
}

\date{Received: March 13, 2022 / Accepted: date}

\maketitle

\begin{abstract}
Many software engineering research papers rely on time-based data (e.g., commit timestamps, issue report creation/update/close dates, release dates).
Like most real-world data however, time-based data is often dirty.  To date, there are no studies that quantify how frequently such data is used by the software engineering research community, or investigate sources of and quantify how often such data is dirty.
Depending on the research task and method used, including such dirty data could affect the research results.
This paper presents an extended survey of papers that utilize time-based data, published in the Mining Software Repositories (MSR) conference series.  Out of the 754 technical track and data papers published in MSR 2004--2021, we saw at least 290 (38\%) papers utilized time-based data.  We also observed that most time-based data used in research papers comes in the form of Git commits, often from GitHub.  Based on those results, we then used the Boa and Software Heritage infrastructures to help identify and quantify several sources of dirty Git timestamp data.  Finally we provide guidelines/best practices for researchers utilizing time-based data from Git repositories.

\keywords{literature review \and time data \and mining software repositories}
\end{abstract}

\section{Introduction}
\label{sec:intro}

The Mining Software Repositories (MSR) conference has been around since 2004 as a workshop, working conference, and finally a full conference.  During those 18 years, there have been over 600 research and over 100 data showcase papers published in MSR proceedings.  The majority of the research in MSR relies on analyzing existing data, including data from version control systems (CVS, Subversion, Git), issue/bug reports, discussion forums (emails, Stack Overflow), pull requests (PRs), continuous build/test systems, etc.  Often these data sources include time components indicating when events occurred, such as the timestamp of a code commit or when a pull request was opened or closed.

Depending on the source of the data, there may be errors or inconsistencies in the time components.  For example, the Git version control system (VCS) allows users to specify both the authored and committed dates when creating new commits. It also allows editing the existing commit graph (rebasing) which allows for modification of the timestamps of older commits.  Similarly, Subversion adds properties to revisions for things like the author and revision date.  While they are added automatically, they can be altered or even removed later. There are also more general issues with time data, for example dealing with inconsistent time zones, clock skews, or more generally, incorrectly set computer clocks.

To date, no survey has been performed to investigate how MSR researchers utilize time-based data in their research.  This work thus surveys 754 MSR technical research and data showcase papers from 2004--2021 to determine how many rely on time-based data and what techniques are utilized to control for potential errors in that data.  We utilize keyword searches of the papers and then manually inspect to determine that at least 209 technical research papers and 81 data showcase papers rely on or provide time-based data.  This accounts for at least 38\% of the papers in MSR's history.  Thus we conclude that time-based data is widely used in MSR research.

Based on the results of the survey indicating that VCS is the most used data kind incorporating time-based data, and that GitHub is the most used data source, we then investigate potential problems with time-based Git data from GitHub.  Since \citet{robles10} previously showed that many research papers in MSR are difficult to reproduce, often due to missing data, we chose to not directly investigate the time-based data used in the prior MSR papers found in the survey.  Instead, we utilized the Boa~\citep{boa,boa-website} and Software Heritage~\citep{software-heritage,software-heritage-archive} infrastructures and attempted to quantify how frequently some types of errors occur in those datasets.  We also attempted to infer the potential source(s) of commonly observed errors.  Based on this investigation, we observed a couple of potential pitfalls when utilizing time-based Git data, and try to quantify how frequently one might encounter such errors if using Git data derived from GitHub.

The results show that almost 50k commits have timestamps that are suspiciously too old (even before the initial release of CVS, 19 November 1990), out of around 54m total commits (around 0.09\%).  Many of those bad timestamps were the result of tools such as \texttt{git-svn}.  We also discovered over 80k commits from over 57k projects where one (or more) of the commit's parent commits had a timestamp that was newer than the commit itself---a situation that does not make sense.  Again, many of these were the result of automated tools or a small set of users.  A replication package containing all of the data and scripts used in our analysis is also publicly available~\citep{replication}.

To help show the potential impact of bad time-based data, we investigate several datasets containing Git repositories from MSR data showcase papers.  These papers collectively have over a hundred citations already.  We intersect the projects in those datasets with the projects from Boa's datasets and find over 15k commits with bad timestamps.  Since those papers already contain over a hundred citations, it shows the potential impact of such bad data propagating to other research and highlights the importance of properly sanitizing data, especially when building reusable datasets.

Finally, we propose some guidelines for researchers utilizing time-based Git data to help escape these pitfalls.  These include filtering out older projects (based on our analysis, we would recommend anything before 2014), filtering out certain projects or users that seem to have a lot of bad commit timestamps, or preferably running specific analyses to automatically verify and reject commits with suspicious timestamps.  We hope future SE researchers follow these guidelines.

Note that this study is an extension of our prior paper~\citep{msrstudy}.  Compared to the previous paper, this work adds the following additional contributions:

\begin{itemize}
    \item We update the survey to include MSR 2021 papers.
    \item We extended the survey analysis to see if there were trends over time in the use of kinds of mined time-based data or data sources.
    \item The original paper analyzed a single dataset consisting of Java projects from GitHub. In this work, we analyze two additional GitHub datasets with Kotlin and Python projects, to see if choice of programming language might influence how often bad time data occurs.
    \item We also analyze a SourceForge dataset from Boa to investigate if CVS and Subversion-based projects from multiple programming languages have bad time data.  This allows us to generalize the results beyond just Git data and beyond the three programming languages studied from GitHub.
    \item For commits that are out-of-order (at least one parent commit is newer than the commit itself), we quantify how far apart the commit is from its out-of-order parent to give some insights into the potential causes.
    \item Finally, we try to quantify the potential impact of such bad time-based data, by analyzing 11 previously published MSR dataset papers and intersecting their data with our datasets to quantify if those datasets contain potentially bad commits.
\end{itemize}

In the next section we discuss related prior research.  In \secref{sec:survey} we detail our survey on the use of time-based data in MSR research.  Then in Sections \ref{sec:approach}--\ref{sec:problems} we attempt to identify and quantify some examples of problems with time-based data in Git/GitHub.  In \secref{sec:impact} we look at the potential impact such time-based data problems might have on the field by analyzing some re-usable datasets.  We discuss implications of the study and present best practice guidelines in \secref{sec:discussion}.  Threats to the validity of our study are then discussed in \secref{sec:threats}.  Finally, we conclude in \secref{sec:conclusion}.

\section{Previous Studies}
\label{sec:related}

In this section we discuss prior works that either performed surveys of MSR research or propose guidelines for future MSR researchers to follow.

\citet{demeyer13} explored 10 years of MSR papers to determine what software projects were studied and the frequency of studies on the given projects, as well as the infrastructure behind mining.  In particular, they noted that the most common source of data were version control systems, including the then-increasing popularity of Git, and infrequency of use of VCSes other than CVS, Subversion or Git.  They also noted that few of the studies at the time had considered industrial cases and instead most were over open source software.  While their work identifies common sources of data (of various kinds), our survey specifically focuses on common sources of time-based data.

\citet{perils,perils2} addressed various characteristics of GitHub repositories. They note several possible problems with GitHub data, such as containing personal and inactive projects or that many pull requests are not marked as merged despite being so. They provide guidelines for software engineering researchers on how to use GitHub data in their research more effectively. Our work is somewhat complementary to theirs.  While they do not focus on time-based issues, that is the focus of our work and our recommendations.

\citet{cosentino16} reviewed the use of GitHub data in prior studies and structured data archives.  In particular, they looked at how GitHub data was used, how the data was collected, and what, if any, limitations were reported in the studies.  The operation of the GitHub API at the time, particularly in terms of request limits and inconsistent responses, was noted as a limitation.  Further, the lack of availability of fresh data was considered as a potential issue, due to reliance on commonly curated data sources.  Finally, they also described potential issues with sampling of datasets, suggesting that better sampling methods are needed.  Similar to their work identifying potential problems with GitHub data, we identify time-based problems with Git data sources and suggest possible methods to avoid such problems when building datasets.

\citet{robles10} was concerned with the replication of MSR studies and observed very few papers were ``replication friendly.''  Replication requires the availability of datasets and tools, as well as an adequate description of techniques used to filter and analyze those datasets.  The tools and descriptions that are preserved for replication may filter using time, yet this particular class of filtering criteria is only one of many which must be considered for replication.  Like \citet{robles10}, \citet{ghezzi13} studied replication of MSR studies.  In particular, they described a web service to gather and analyse software repository data, which was then used to replicate MSR studies from 2004--2011.  They found that, of the studies in those years, 51\% could not be fully replicated.  While these works looked at prior papers to estimate replicability, our work looks at prior papers to investigate the use of time-based data.  We also rely on this work to help motivate portions of our study.

\citet{kotti19} investigated the use and impact of datasets published as MSR data showcase papers via a systematic review.  They noted that a number of further work has built upon the MSR data showcase papers, with over 65\% being used in other studies and one having 157 (at time of their publication) citations.  In particular, they clearly show the impact of the data showcase papers, however, they do not investigate potential issues with the re-use of this data or possible problems with the datasets.  In our work, we look at potential problems with some of these datasets.  We rely on their results to motivate that investigation.

\citet{hemmati13} described a set of best-practices, a ``cookbook'', for mining software repositories researchers.  This included suggestions regarding social data, statistical and analytical techniques, and the sharing of tools.  They discuss the issue of VCS noise and the potential lack of granularity in VCS-recorded changes, however, they do not discuss the potential causes of discontinuities in time data, nor ways they may be handled.  In this work we not only identify several examples of problems with time-based data, but also propose some guidelines on how to clean and filter data to avoid those problems.

\citet{gasser04}, early in MSR's history, evaluated the needs of researchers in the field and the data and artifacts to be studied.  They proposed a set of characteristics for studies to have, and discussed issues with data and how these issues may be addressed.  In particular, they discussed the frequent need to \emph{normalize} data as part of the analysis and data collection process.  They did not however focus on time-based data, which is the focus of this study.

\citet{bird09} discussed mining one VCS in particular, Git, and the potential issues that may occur in mining repositories using it.  This work describes a number of issues, in particular, the existence of rebasing, which allows users to rewrite a repository's history, re-using commits in a different order than the commit timestamps may suggest.  Thus their work helps identify potential sources of bad time-based data, while in our work we propose some guidelines to avoid including bad time-based Git data.

Many of the prior works mentioned here either performed surveys to look at prior studies or they discuss common issues studies face and provide guidelines.  These works, unlike our study, were not focused on time-based data, or on providing guidance on how to deal with potentially bad time-based data.

\section{Survey on the Use of Time-Based Data}
\label{sec:survey}

This work first investigates the following research question:

\begin{enumerate}[label={\textbf{RQ\arabic*}},leftmargin=2.5\parindent]
    \item \label{rq:study} \textbf{Does prior software mining research rely on time-based data?} We focus on the mining software repositories (MSR) conference series, as it is the preeminent conference for mining repositories.  Thus we investigate: is time-based data commonly utilized in published MSR technical and data showcase papers?
\end{enumerate}

\noindent Based on the results of this research question showing a large number of papers using time-based data, we then investigate two other questions by analyzing the subset of papers that utilize time-based data:

\begin{enumerate}[label={\textbf{RQ\arabic*}},leftmargin=2.5\parindent,start=2]
    \item \label{rq:kinds} \textbf{What kinds of commonly mined software data include time?} When looking for time-based data, are there common kinds and sources for that data? This can give insight into what problems might occur and where researchers should focus their efforts to ensure the data is properly sanitized.

    \item \label{rq:filtering} \textbf{What filtering or cleaning techniques are used with time-based data?} Do papers already use filtering or cleaning techniques for time-based data, and if so are such techniques common? Can we infer recommendations based on the existing approaches?
\end{enumerate}

We begin by surveying published MSR proceedings.  We select papers to review, then from these, classify what kind of time-related data is used, how it is filtered or cleaned in published work, and the sources of time-based data.

\subsection{\ref{rq:study}: Does prior software mining research rely on time-based data?}
\label{subsec:paper-selection}

For this study we focused only on papers published in MSR proceedings from 2004 to 2021.  All technical track papers (short and long) and Data Showcase papers were considered.  Data Showcase papers were included as they are potential data sources for other (future) research papers.  Mining Challenge papers were excluded, as all papers in this category for a given year typically use the same challenge dataset, which may skew results towards a particular kind of data in that year.  This gave us a corpus of 754 papers to inspect.

One author used a keyword search to filter papers from this corpus, retaining for further study those papers that contained any of the following time-related keywords\footnote{All authors brainstormed potential keywords and helped create the final list.}:  \texttt{time}, \texttt{date}, \texttt{epoch}, \texttt{record}, \texttt{month}, \texttt{year}, \texttt{hour}, \texttt{minute}, \texttt{second}, \texttt{period}, \texttt{week}, \texttt{chronolog}, \texttt{day}, \texttt{past}, and \texttt{interval}.  This retained 346 of the 754 papers (45.89\%).

After papers were initially filtered, two authors independently analyzed each paper to determine what kinds of time-based data were used, the source(s) of the data, and any methods used to filter, clean or normalize the time-based data.  During this process, if either author found that a paper did not fit the study, it was voted for removal and removed if a second author agreed.  For example, papers which used ``runtime'' as a performance metric, or ``epoch'' as a measure of training time were considered as irrelevant to the study and removed if no other time-based data was used.

After the first round, if the two authors disagreed on the kinds of data, data sources, or filtering techniques, they discussed this disagreement until consensus was reached.  This affected a total of 16 papers (for source of data), 12 papers (for kinds of data), and 13 papers (for filtering techniques) for a total of 38 papers (3 papers had several disagreements).  This data is described in more detail in the following subsections.

\begin{table*}
\centering
\caption{Published and selected MSR papers, by year. The full list of considered papers and the human judgements made by the authors are available in an Excel spreadsheet in our replication package~\citep{replication}. The full list of selected papers is also shown in \appref{app:papers}.}
\label{tab:years}
\begin{tabular}{rr@{\hskip.2em}!{/}@{\hskip.2em}lr@{\hskip.2em}!{/}@{\hskip.2em}lc}
  \toprule
                 & \multicolumn{2}{c}{\textbf{Technical Papers}} & \multicolumn{2}{c}{\textbf{Data Showcase}} & \textbf{Percent}                                                                 \\
                 & \textbf{selected}                      & \textbf{total}                    & \textbf{selected}    & \textbf{total} & \textbf{Selected} \\
  \midrule
  2004           & 5                                      & 26                                & \multicolumn{2}{c}{} & 19\%                                       \\
  2005           & 2                                      & 22                                & \multicolumn{2}{c}{} & 9\%                                        \\
  2006           & 6                                      & 28\(^{*}\)                        & \multicolumn{2}{c}{} & 21\%                                       \\
  2007           & 8                                      & 27                                & \multicolumn{2}{c}{} & 30\%                                       \\
  2008           & 7                                      & 31                                & \multicolumn{2}{c}{} & 23\%                                       \\
  2009           & 9                                      & 23                                & \multicolumn{2}{c}{} & 39\%                                       \\
  2010           & 6                                      & 22                                & \multicolumn{2}{c}{} & 27\%                                       \\
  2011           & 9                                      & 27                                & \multicolumn{2}{c}{} & 33\%                                       \\
  2012           & 9                                      & 31                                & \multicolumn{2}{c}{} & 29\%                                       \\
  \midrule
  2013           & 10                                     & 37                                & 9                    & 14             & 37\%                      \\
  2014           & 13                                     & 39                                & 12                   & 15             & 46\%                      \\
  2015           & 12                                     & 42                                & 13                   & 16             & 43\%                      \\
  2016           & 11                                     & 42\(^{\dagger}\)                  & 4                    & 7              & 31\%                      \\
  2017           & 13                                     & 43                                & 6                    & 7              & 38\%                      \\
  2018           & 13                                     & 48                                & 10                   & 15             & 37\%                      \\
  2019           & 16                                     & 47                                & 2                    & 17             & 28\%                      \\
  2020           & 20                                     & 45                                & 14                   & 19             & 53\%                      \\
  2021           & 40                                     & 48                                & 11                   & 16             & 80\%                      \\
  \midrule
  \textbf{Total} & \textbf{209}                           & \textbf{628}                      & \textbf{81}          & \textbf{126}   & \textbf{38\%}             \\
  \bottomrule
\end{tabular}
\\
\centering
{\scriptsize
($^*$2006 had 2 papers listed in the program that do not appear in the proceedings, which were excluded)\\
($^{\dagger}$2016 had 1 paper not listed in the program that appears in the proceedings, it was included)}

\end{table*}

A total of 56 papers (7.4\%) had a matching keyword but were removed, leaving 290 papers.  Thus the simple keyword search yielded a precision of 84\%.  The results of this selection process are shown in \tabref{tab:years} and the spreadsheet including all considered and selected papers and human judgements made by the authors is available in the replication package~\citep{replication}. A full citation of all selected papers is available in \appref{app:papers}.

The results show that every year of MSR had papers relying on time-based data, ranging from 9-80\% of all papers in a given year.  Both the technical and data showcase tracks have papers in every year relying on time-based data.

\findings{
    Time-based data is prevalent in MSR research papers, accounting for anywhere from 9-80\% of the papers in a given year.  On average, 38\% of all MSR papers utilize time-based data.
}

\subsection{\ref{rq:kinds}: What kinds of commonly mined software data include time?}
\label{subsec:classification}

Since so many papers utilize time-based data, we now investigate if there are common kinds and sources of time-based data.  We found that across the 290 papers selected, 37 different kinds of time-including data were used.  From these, all data kinds used by more than one paper are shown in \tabref{tab:data-kinds}.  In particular, we found that VCS data (diffs, commit lineage, commit logs, authors, etc.) are the most commonly used.  In addition, we found that issues and their related metadata were used frequently as well.

We keep some similar kinds, such as Mailing List, Developer Q\&A (e.g., StackExchange, StackOverflow), and Chat Logs separate as although each serve a somewhat similar purpose, researchers tend to ask different questions about them.  We kept several categories of logs separate (e.g., ``General Logs'' from logging frameworks or servers, user interaction logs, chat logs) for similar reasons.

\begin{table}
\caption{Common kinds of data used in MSR papers. Only data kinds used by more than one paper are listed here.}
\label{tab:data-kinds}
\centering
\begin{tabular}{r@{\hskip.5em}ll}
  \toprule
\multicolumn{2}{c}{\textbf{Num. Papers}} & \textbf{Data Kind} \\ 
  \midrule
188 & (64.83\%) & VCS \\ 
  81 & (27.93\%) & Issues \\ 
  46 & (15.86\%) & Releases \\ 
  42 & (14.48\%) & Forge Metadata \\ 
  19 & (6.55\%) & Mailing List \\ 
  19 & (6.55\%) & Pull Requests \\ 
  18 & (6.21\%) & Developer Q\&A \\ 
  13 & (4.48\%) & Continuous Improvement Logs \\ 
  13 & (4.48\%) & General Logs \\ 
  7 & (2.41\%) & Interaction Logs \\ 
  6 & (2.07\%) & Common Vulnerabilities and Exposures (CVEs) \\ 
  3 & (1.03\%) & Time Cards \\ 
  2 & (0.69\%) & Chat Logs \\ 
  2 & (0.69\%) & File Dates \\ 
   \bottomrule
\end{tabular}

\end{table}

\begin{figure}[htbp]
  \centering
  \includegraphics[width=0.99\linewidth]{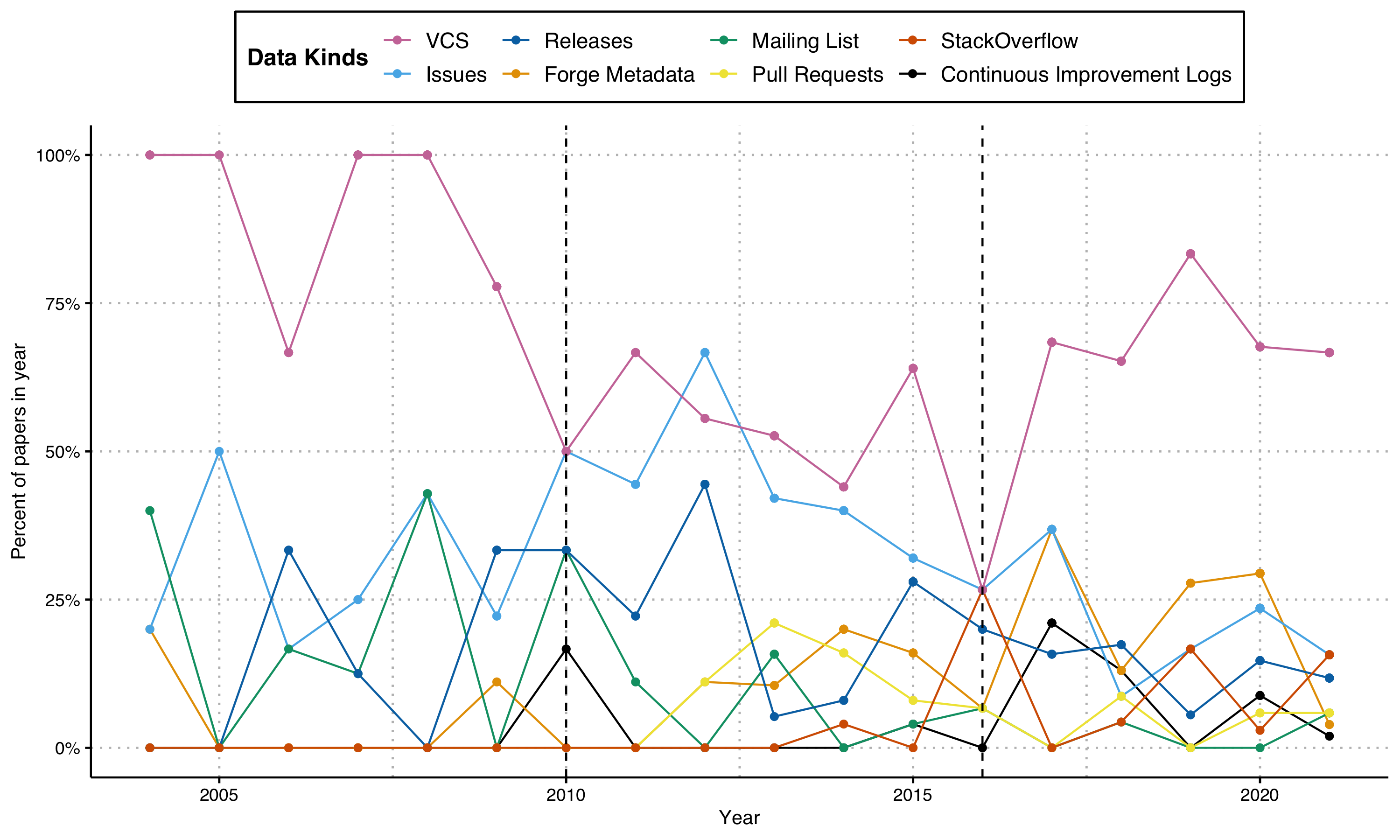}
  \caption{Kinds of data used over time (top 8). Each data point represents the percent of papers that year that utilized that kind of data.}
  \label{fig:data-kinds-over-time}
\end{figure}

We also investigated if the kinds of data used in MSR papers has changed over time.  We show a graph of the eight most common data kinds in \figref{fig:data-kinds-over-time}.  This graph has time on the x-axis (in years) and for each year, then plots the percent of papers utilizing that particular data kind.  Note that the totals may add up to more than 100\% as some papers utilize multiple data kinds.

When visually inspecting the graph, several trends emerge.  We identified three different time periods with different trends. The first period is from 2004--2010, where a large percent of papers utilize VCS data, but there was also several other data kinds frequently used such as issue data and mailing list data.

The second period was from 2010--2016.  During this period, VCS data use (as a percentage) dropped down.  Now we were starting to see more papers that focused on a specific, non-VCS form of data such as issues, releases, or pull requests.  In particular, issue data was quite popular (about as popular as VCS data) during this period.

The third period is from 2016--2021.  During this period, we can see a clear shift back toward a focus on VCS data.  While other forms of data still appear, a large percentage (60\%+) of papers every year rely on VCS data while other forms of data only appear in about a quarter or less of papers.  The popularity of data kinds such as issues decreased.  This helps show the importance of having good VCS data, as bad data may affect many future papers.

\findings{
    Except for a period from around 2010--2016, VCS data seems to be the most popular kind of time-including data used by MSR researchers.
}

We also investigate the sources of data, that is, where the data is gathered from (as opposed to what the data is) and found roughly 209 different sources were used, with proprietary and anonymous repositories listed as a single, ``anonymized'' source.  We then categorized these sources to provide a higher-level overview of some of the most common and to group them together by their (perceived) similarity to each other.  These categories and some of the more common sources within the category are shown in \tabref{tab:data-sources}.  Perhaps unsurprisingly, GitHub is the most common \emph{single} data source, but FOSS project repositories, such as Eclipse, Apache, and Mozilla, are also quite common.  It is also notable that the GHTorrent~\citepappD{6224294} dataset is frequently used as well.

\begin{table}
\caption{Common data sources used by MSR papers. Only data sources used by more than one paper are listed here. Top individual data sources are indicated by bold.}
\label{tab:data-sources}
\centering
\begin{tabularx}{0.95\linewidth}{llX}
  \toprule
  \textbf{Source}      & \textbf{Total}  &                                                                                                                                                        \\
  \textbf{Category}    & \textbf{Number} & \textbf{Examples}                                                                                                                                      \\
  \midrule
  FOSS projects        & 103 (35.52\%)   & \textbf{Eclipse (23, 7.93\%)}, \textbf{Apache (19, 6.55\%)}, Mozilla, Firefox, PostgreSQL, OpenStack, ArgoUML, GCC, Python, Chrome                     \\
  forges               & 85 (29.31\%)    & \textbf{GitHub (81, 27.93\%)}, \textbf{GHTorrent (18, 6.21\%)}, GitLab, SourceForge, BitBucket, Software Heritage Archive, DockerHub, Google Code, Boa \\ 
  anonymized           & 24 (8.28\%)     & various                                                                                                                                                \\ 
  operating systems    & 23 (7.93\%)     & Linux Kernel, RedHat, Debian, Gentoo, Fedora, BSD                                                                                                      \\ 
  social               & 21 (7.24\%)     & \textbf{Stack Overflow (16, 5.52\%)}, Twitter, Devpost                                                                                                 \\ 
  app stores           & 9 (3.10\%)      & Google Play, F-Droid                                                                                                                                   \\ 
  issue trackers       & 8 (2.76\%)      & Jira, Gerrit, BugZilla                                                                                                                                 \\ 
  CI systems           & 6 (2.07\%)      & Travis, Jenkins, TeamCity                                                                                                                              \\ 
  package repositories & 6 (2.07\%)      & Maven Central Repository, Comprehensive R Archive Network, PyPi, Comprehensive Perl Archive Network, Node Package Manager                              \\ 
  security related     & 5 (1.72\%)      & National Vulnerability Database, Common Platform Enumeration, CAPE, Common Vulnerabilities and Exposures                                               \\ 
  messaging            & 2 (0.69\%)      & Slack, Gitter                                                                                                                                          \\ 
  \bottomrule
\end{tabularx}
\end{table}

To help show overlap between sources and data kinds, we present in \tabref{tab:source-kind-pairs} the top 10 source/kind pairs, and the percent the pair is seen in the studied papers as well as in all MSR published papers.  In particular, we note that around 23\% of the papers studied (or about 9\% of published MSR papers) use time-based VCS data and specifically get data from GitHub.  Forge metadata from GitHub is also common at about 3\% of all MSR papers, with the combination of Issues and either Eclipse or GitHub sources the third-most common combination.

\begin{table}
\centering
\caption{Top 10 data source/data kind pairs.}
\label{tab:source-kind-pairs}
\begin{tabular}{llll}
  \toprule
                      & \textbf{VCS} & \textbf{Forge Metadata} & \textbf{Issues} \\ 
  \cmidrule{2-4}
  \textbf{GitHub}     & 22.76\%      & 8.97\%                  & 6.21\%          \\ 
  \textbf{Eclipse}    & ~4.83\%      &                         & 6.21\%          \\ 
  \textbf{GHTorrent}  & ~5.52\%      & 3.79\%                  &                 \\ 
  \textbf{anonymized} & ~5.17\%      &                         & 3.10\%          \\ 
  \textbf{Apache}     & ~3.45\%      &                         &                 \\ 
  \bottomrule
\end{tabular}

\end{table}

\begin{figure}[htbp]
  \centering
  \includegraphics[width=0.99\linewidth]{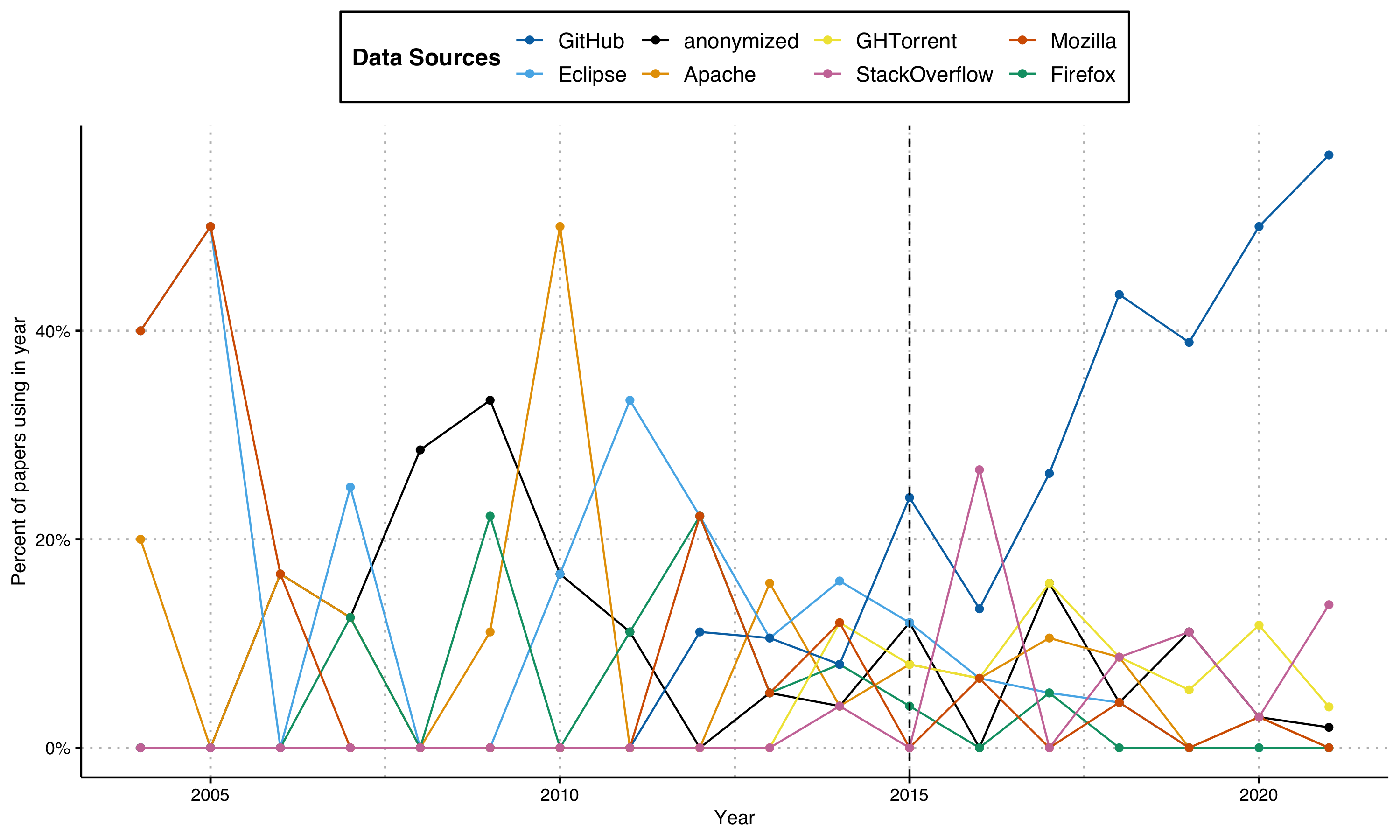}
  \caption{Data sources used over time (top 8). Each data point represents the percent of papers that year that utilized that data source.}
  \label{fig:data-sources-over-time}
\end{figure}

Similar to the analysis of data kinds over time, we visually analyzed the data sources over time, shown in \figref{fig:data-sources-over-time}.  Here the data was a bit noisier, but the data falls roughly into two time periods: before and after 2015.

Before 2015, the sources of data were much larger than after 2015.  There did not seem to be a consensus on which dataset(s) to use for MSR papers, as things like the Apache repositories or the Mozilla repositories saw huge swings from 40\%+ use in one year to 0\% a year or two later.  Notably, GitHub did not exist for half of that period and did not really start gaining in popularity until around 2012.

The second period, after 2015, is where we see GitHub's popularity starting to affect the choice of data MSR researchers used.  You can see a steady incline in the percent of MSR papers using GitHub as a (direct) data source.  While other sources are still used, they typically account for a small subset of papers in any given year.  Note here that when we graph GitHub, we are graphing direct uses of GitHub data -- not indirect uses through aggregated data such as GHTorrent~\citepappD{6224294} (shown separately).  This shows that MSR researchers are greatly preferring to use data from GitHub, and thus we need to understand what issues may be present in that data.

\findings{
    It appears many popular sources in the past, such as Eclipse and Apache, are seeing declining use while GitHub is becoming the clear leading source of time-based data.
}

\subsection{\ref{rq:filtering}: What filtering or cleaning techniques are used with time-based data?}
\label{sec:filtering}

Finally we investigated any filtering or cleaning techniques used by the selected papers.  Our goal was to see if there are any commonly used methods, so that we can inform future researchers of such techniques.  We identified 54 different methods of cleaning or filtering the time-based data (considering all custom conditions as one method for the purposes of counting).  We show any used by more than one paper in \tabref{tab:filtering-methods}.

\begin{table}
\caption{Common filtering/cleaning techniques used by MSR papers. Only techniques used by more than one paper are listed here.}
\label{tab:filtering-methods}
\centering
\begin{tabular}{r@{\hspace{0.2em}}ll}
  \toprule
\multicolumn{2}{c}{\textbf{Number of Papers}} & \textbf{Filtering Technique} \\
  \midrule
 192 & (66.21\%) & none explicitly mentioned \\
  30 & (10.35\%) & time window \\
  24 &  (8.28\%) & date cutoff \\
  12 &  (4.14\%) & custom condition \\
   5 &  (1.72\%) & changeset coalesence \\
   5 &  (1.72\%) & date format correction \\
   4 &  (1.38\%) & CVSAnalY \\
   \bottomrule
\end{tabular}

\end{table}

Among the various time-based filtering and cleaning techniques found in MSR papers, we found six used by more than one paper.  The majority of these are filtering techniques of some form, with a single cleaning technique described.  It is important to note, however, the majority of papers utilizing time-based data (192, 66.21\%) do not explicitly describe any filtering or cleaning methods used, although they might have performed some filtering or cleaning.  We discuss each of the six mentioned techniques in more detail here.

\subsubsection{Time Window}

A number of studies select data from a source that was added between two dates or other markers in time (e.g., releases).  This was by far the most common explicitly described method, being found in 30 studies.  Some of these studies provided full dates~\citep{durieux20,pimentel19}, others only partial dates out to year or month~\citep{hayashi19,ahasanuzzaman16}, or version numbers of releases~\citep{antoniol05}.

\subsubsection{Date Cutoff}

All studies retrieved data from before or after a specific date (the date of the study).  But whether the study date is used or some other date is used must be considered.  In particular, some papers describe what their cutoff date for data inclusion is, while others do not.  This method is used in particular by~\citet{Wang_2020,Karampatsis_2020,Zhu_2019,Cito_2017}.

\subsubsection{Custom Condition}

A custom condition specifies some method for filtering a data source using time.  These were frequently employed to ensure that commits or issues were studied that matched some temporal condition relating the two, or to ensure that commits were in order, as well as for other purposes.

\citet{liu2020} describe the use of a particular time-based condition to select commits to study.  They were interested in finding commits between the open and close of a particular issue (in other words, looking for fixing commits).  This condition is $issue_{create} < commit_{create} < issue_{close}$, and uses time components from both issues and commits.

\citet{kikas16} use commit time and a forge's repository creation time to remove forks of original projects so that only the originals may be studied.  We note in particular that this method may inappropriately remove projects which have changed forges.

Finally, \citet{steff12} construct a commit graph such that, for each commit node, it is only connected to nodes preceding it in time which also share files in common, that is, for two commits $(t_1, \mathcal{F}_1)$ and $(t_2, \mathcal{F}_2)$, $(t_1, \mathcal{F}_1) \to (t_2, \mathcal{F}_2)$ if and only if $t_1 < t_2$ and $\mathcal{F}_1 \cap \mathcal{F}_2 \neq \emptyset$.

\subsubsection{Changeset Coalescence}

Further techniques used include changeset coalescence or commit reconstruction.  This technique is useful in CVS or RCS repositories where changes are only made to individual files.  Most of these methods operate by collecting changes made in a small window (3 minute) by one user into a single changeset; they may also be aided by the use of ChangeLogs to collect such changes.  This technique was used by~\citet{zimmerman04,walker06,kagdi06,dambros10}.

\subsubsection{CVSAnalY}

CVSAnalY\footnote{\url{https://github.com/MetricsGrimoire/CVSAnalY}} is a tool to extract information from VCS logs of various kinds.  It supports CVS, Subversion, and Git.  When operating on Subversion repositories, it skips over commits it considers invalid, with one condition being the lack of date\footnote{Note that: ``While Subversion automatically attaches properties (svn:date, svn:author, svn:log, and so on) to revisions, it does not presume thereafter the existence of those properties, and neither should you or the tools you use to interact with your repository.'' \url{https://svnbook.red-bean.com/en/1.7/svn.advanced.props.html}}.  Otherwise, it performs a sort of date format correction, storing all dates as Unix timestamps with associated time zones.

In particular, this tool sees use on Git repositories~\citep{gonzalez-barahona15,robles14,goeminne13}, as well as Subversion repositories~\citep{sadowski11}, where the filtering may be most apparent.

\subsubsection{Date Format Normalization}

Due to the diversity of data sources and systems used, date and time data often must be normalized, that is, put into a standard format.  This may include time zone conversion or other actions, and presents a single, unified view of time for analysis and further filtering.  This process makes it easier to further process since all data will be in the same format and thus is considered a cleaning technique.  As an example, it is specifically used by~\citet{claes20,xu18,baysal12}.

These are some of the most common time-based data filtering techniques used.  In the next section we investigate and attempt to quantify how frequently problems occur in time-based data.

\subsubsection{Summary}

33.79\% of papers using time-based data describe using some sort of filtering or cleaning method: mining from within specific windows, mining from before a specific date, or using date format correction.  Twelve papers describe using custom, specialized conditions or methods based on the needs of the study.   Further, the papers that use CVS or RCS describe using a method to coalesce changes into changesets, and several papers used CVSAnalY to perform their mining.

\findings{
    Filtering and cleaning techniques are used in less than 34\% of MSR papers using time data.  Some of the more generally applicable techniques used include filtering to a time window, using a specific date cutoff, or normalizing dates to a standard format.
}

\section{The Pitfalls of Time-Based VCS Data}
\label{sec:approach}

In the last section, we surveyed the MSR literature to determine how often time-based data is used.  Our results showed that time-based data is frequently used--by at least 38\% of MSR papers surveyed.  We then looked at what the most frequently used data kinds and data sources were.  Based on those results, we observed that VCS data (often Git) is the most used data kind and, specifically, data from GitHub tends to be the most used data source.  Based on these results, we identified an additional research question to investigate:

\begin{enumerate}[label={\textbf{RQ\arabic*}},leftmargin=2.5\parindent,start=4]
    \item \label{rq:bad-data} \textbf{How frequently does \textit{bad} time-based VCS data occur?} Based on the results of the prior research questions, we investigate Git data from GitHub to quantify how frequently bad time-based data occurs.  Additionally, we investigate Subversion/CVS data from SourceForge to show that these problems may exist in data from sources other than Git.
\end{enumerate}

Since the previous section showed a large number of MSR research papers rely on time-based data, it is important to get a feel for how often such data might be bad.  To date, no study has investigated how frequently bad time-based data occurs and thus we do not know if existing (or future) studies are relying on a large amount of possibly bad data.

In the remainder of this section, we outline our methodology.  We do not directly analyze the papers studied in the prior section, as prior research showed that many research papers are not possible to replicate, often because of missing data~\citep{robles10,ghezzi13}.  Given this fact, we instead opt to directly analyze a large number of open-source repositories and see if we can quantify how often bad time-based data occurs in the wild.  In a later section, we then investigate how often those repositories with bad data are used by some MSR data papers.

\subsection{Datasets Studied}

To attempt to quantify problems with time-based data, we utilize the Boa infrastructure~\citep{boa,boa-website}.  We use Boa as it provides many large, diverse datasets which have been pre-processed to ease analysis and we are already familiar with it.  Because Boa provides these large, pre-processed datasets, we are not required to re-collect and build a dataset, speeding our analyses and enabling easier replication.  We rely on four different datasets: ``2013 September/SourceForge'', ``2019 October/GitHub'' (containing Java project commits), ``2021 Aug/Kotlin'', and ``2021 Aug/Python''.  An overview of these datasets is shown in \tabref{tab:dataset-characterestics}, listing the total number of projects and commits in each dataset.

\begin{table}[ht]
  \centering
  \caption{Summary of dataset characteristics.}
  \label{tab:dataset-characterestics}
  \begin{tabular}{lrrrr}
  \toprule
                                    & \textbf{SourceForge} & \textbf{Java}  & \textbf{Kotlin} & \textbf{Python} \\
                                    & (SVN) & (Git) & (Git) & (Git) \\
  \midrule
  \textbf{Projects}                 & 65,934               & 282,781        & 499,645         & 100,940          \\
  \textbf{Commits}                  & 15,063,073        & 23,229,406     & 11,022,118       & 5,427,215       \\
  \bottomrule
\end{tabular}

\end{table}

The first of these is composed of open-source projects from SourceForge, the remaining three datasets contain open-source projects from GitHub.  The SourceForge dataset contains around 65k Subversion (SVN) projects with at least one revision from over 50 programming languages, including over 23k Java projects, 9k C++ projects, 4k C projects, 4k C\# projects, and 900 PHP projects. The Java dataset contains over 200k projects, the Kotlin dataset contains almost 500k projects~\footnote{The Kotlin dataset contains some projects which may exist in the Java dataset.}, and the Python dataset contains around 100k projects. In total, these four datasets provide over 54 million total commits.

For some of the analyses, we needed to know a commit's list of parent commit hashes.  Some of the older Boa datasets did not directly provide that information, so we utilize the GitHub API to obtain it.  Since some of the repositories in the Boa dataset also no longer exist on GitHub, we would not be able to utilize their API to obtain that information.  As such, we also utilized the Software Heritage Archive~\citep{software-heritage,software-heritage-archive} to attempt to locate any repositories in the Boa datasets that were deleted from GitHub since the Boa dataset was built.

We identify and remove exact duplicate commits (same commit hash) from the GitHub datasets\footnote{The SF.net dataset contained Subversion projects, which store commit IDs as integers and thus are not unique across projects and can not be easily deduplicated.}.  All analyses in this paper utilize the de-duplicated datasets.

\subsection{Query Approach}

\figref{fig:boaquery} shows the relevant Boa query used to collect data for our investigation.  Note that Boa stores commits in a topologically sorted array based on the commit parent(s).  This means traversals on the commits (called \texttt{Revision}s in Boa) are performed in topological order.

\begin{figure}[ht]
\begin{Boa}
P:  output collection[string][string] of time;
P2: output collection[string][string] of time;
P3: output collection[string][string] of time;

cvs_release_date := T"Mon Nov 19 00:00:00 UTC 1990";
boa_dataset_date := T"Thu Oct 31 00:00:00 UTC 2019";%\label{ln:date}%
last: Revision;

visit(input, visitor {
	before r: Revision -> {
		if (r.commit_date < cvs_release_date)%\label{ln:old1}%
			P[input.project_url][r.id] << r.commit_date;%\label{ln:old2}%
		else if (r.commit_date > boa_dataset_date)%\label{ln:future1}%
			P2[input.project_url][r.id] << r.commit_date;%\label{ln:future2}%

		if (def(last)%\label{ln:order1}%
		    && r.commit_date < last.commit_date
		    && !match("merge", lowercase(last.log))
		    && !match("merge", lowercase(r.log)))
			P3[input.project_url][r.id] << r.commit_date;
		last = r;%\label{ln:order2}%
	}
});
\end{Boa}
  \caption{Boa query to find bad commit timestamps in the Java dataset.  This query is the combination of \url{http://boa.cs.iastate.edu/boa/?q=boa/job/public/90164}, \url{http://boa.cs.iastate.edu/boa/?q=boa/job/public/90169}, and \url{http://boa.cs.iastate.edu/boa/?q=boa/job/public/90973} for presentation purposes.  Similar queries were built for the other datasets by changing the date on line \ref{ln:date}.}
  \label{fig:boaquery}
\end{figure}

The query outputs the project URL, the commit ID, and any suspicious commit timestamp.  The query looks for three possible kinds of bad time data indexed by project URL and revision ID.  First, it looks for suspicious commit timestamps that seem too old (lines \ref{ln:old1}--\ref{ln:old2}).  We define a commit as being suspiciously old if it occurred prior to the release date of CVS (Nov 19, 1990), as our datasets are based on SVN and Git and post-date even CVS's release.

Second, it looks for suspicious commit timestamps that seem too new (lines \ref{ln:future1}--\ref{ln:future2}).  Here we use the date of the dataset itself as the definition of the current time, and look for any commit in the ``future.''  This date is changed for each dataset to match the date of the dataset.

Third, it looks for commits that have a parent that is newer than themselves (lines \ref{ln:order1}--\ref{ln:order2}).  Since the commits are ordered topologically based on time, a commit should always have a timestamp that is not older than its parent's timestamp.

In the next section, we investigate how frequently bad time-based VCS data occurs in the studied datasets.

\section{\ref{rq:bad-data}: How frequently does bad time-based VCS data occur?}
\label{sec:problems}

In this section we investigate some potential problems with Git, SVN, and CVS timestamps and attempt to quantify how frequently such problems occur in the wild.

\subsection{Looking for Suspicious Commit Timestamps}
\label{subsec:old-commits}

First we investigated to see if there were suspicious commit timestamps within the studied repositories.  For Git repositories, one might expect the commit timestamps to be after the initial release of Git (around 2005).  It is however possible some repositories were in a different version control system (such as CVS or Subversion) and converted to Git.  For the sake of this study, we decided to investigate any commit timestamp prior to the release of CVS version 1.0 (19 November 1990).  The relevant part of the Boa query in \figref{fig:boaquery} is lines \ref{ln:old1}--\ref{ln:old2}.

The result of this query found 3,612 suspicious commit timestamps from 51 projects in the Java dataset.  For those projects, this represents 4.20\% of their total commits.  For the full dataset, this represents 0.02\% of the commits.  In total, there were 23 unique suspicious timestamps (note: Boa timestamps are given as Unix timestamps with milliseconds), listed in \tabref{tab:oldcommits} along with the number of times they occurred and their conversion to a human readable date format.

\begin{table}
    \centering
    \caption{Suspiciously old commit timestamps in the Java dataset}
    \label{tab:oldcommits}
    \begin{tabular}{rrl}
  \toprule
\textbf{Count} & \textbf{Timestamp} & \textbf{Date/Time} \\
  \midrule
   1 & -2044178335000000 & 03/23/1905, 12:41:05 PM \\
3576 & 0                 & 01/01/1970, 12:00:00 AM \\
   1 & 730000000         & 01/01/1970, 12:12:10 AM \\
   1 & 956000000         & 01/01/1970, 12:15:56 AM \\
   1 & 1585000000        & 01/01/1970, 12:26:25 AM \\
   1 & 1601000000        & 01/01/1970, 12:26:41 AM \\
   1 & 1627000000        & 01/01/1970, 12:27:07 AM \\
   1 & 3495000000        & 01/01/1970, 12:58:15 AM \\
   1 & 3523000000        & 01/01/1970, 12:58:43 AM \\
   1 & 7403000000        & 01/01/1970, 02:03:23 AM \\
   1 & 7558000000        & 01/01/1970, 02:05:58 AM \\
   1 & 7923000000        & 01/01/1970, 02:12:03 AM \\
   1 & 88210000000       & 01/02/1970, 12:30:10 AM \\
   2 & 88211000000       & 01/02/1970, 12:30:11 AM \\
   3 & 88212000000       & 01/02/1970, 12:30:12 AM \\
   2 & 88213000000       & 01/02/1970, 12:30:13 AM \\
   1 & 127771000000      & 01/02/1970, 11:29:31 AM \\
   1 & 179895000000      & 01/03/1970, 01:58:15 AM \\
   1 & 255447000000      & 01/03/1970, 10:57:27 PM \\
  11 & 1000000000000     & 01/12/1970, 13:46:40 PM \\
   1 & 315772873000000   & 01/03/1980, 06:41:13 PM \\
   1 & 566635987000000   & 12/16/1987, 06:53:07 AM \\
   1 & 589770257000000   & 09/09/1988, 02:04:17 AM \\
  \bottomrule
\end{tabular}

\end{table}

SourceForge had a similar large number of older commits (46,266).  For that data, the majority of them were the timestamp \texttt{-1} and there were no commits with timestamp \texttt{0}.  Kotlin and Python however showed different results.  Kotlin only had 6 old commits and Python only had 20 old commits.  This might be due to the fact these are newer languages (Kotlin) or recently becoming popular (Python) and thus many projects might have started directly with Git.

As can be seen from the Java results above, the majority of the suspicious timestamps are the value 0.  There are however a handful of other suspicious timestamps.  For example, the 8 timestamps on January 2, 1970 at 12:30 all come from a single project that was ported over from Microsoft's CodePlex.\footnote{\url{https://archive.softwareheritage.org/browse/origin/log/?origin_url=https://github.com/KevinHoward/Irony&timestamp=2015-07-29T09:07:18Z}}  Most likely there was a problem in that porting process.

In fact, many of these suspicious timestamped commits seem to come from tools, such as \texttt{git-svn}.\footnote{\url{https://archive.softwareheritage.org/browse/origin/log/?origin_url=https://github.com/maodouzi/PY&timestamp=2015-08-07T07:29:54Z}}  This tool was popular in the period between when Subversion was more common and people were starting to move to Git.  It allows maintaining a Git clone of a Subversion repository but required inserting `git-svn-id' tags into the commit messages to properly track the SVN repository.  We were able to verify 3,153 of the commit logs (for Java, Kotlin, and Python) via GitHub's API, and 2,847 of those commits (90\%) contain a \texttt{git-svn-id} tag in the message.

\begin{table}
\centering
\caption{Table of frequent tokens appearing in suspicious commit messages, excluding commits containing the frequent term ``git-svn-id''.  English stop words were removed.}
\label{tab:wordcloud-old}
\begin{tabular}{r@{\hspace{0.5em}}l@{\hspace{10em}}r@{\hspace{0.5em}}l}
  \toprule
  \textbf{\(n\)} & \textbf{Token} & \textbf{\(n\)} & \textbf{Token} \\
  \midrule
309 & process & 10 & enter    \\
309 & http    & 10 & rc       \\
18  & ad      & 7  & empti    \\
18  & add     & 6  & check    \\
14  & merg    & 6  & make     \\
13  & move    & 6  & initi    \\
13  & fix     & 6  & thi      \\
12  & commit  & 6  & minor    \\
12  & bug     & 6  & function \\
10  & git     & 6  & chang    \\
\bottomrule
\end{tabular}

\end{table}

Since that tool accounted for such a large portion of the commits, we investigated all the remaining commits, to look for other possible common tools, by generating a table of frequently occurring words in the commit logs.  We generated this table by using NLTK\footnote{\url{https://www.nltk.org/}} to first tokenize the commit messages, then removing all English stop words, and lemmatizing and stemming the remaining words.  We then group and sort the remaining tokens, shown in \tabref{tab:wordcloud-old}.  Note that we also removed any commits containing ``git-svn-id''.  What remains does not seem to indicate any additional tools accounting for a large portion of the commits.

We also investigated dates that might be in the future. For this we used a cutoff time of the Boa dataset's release date (e.g., 31 October 2019 for the Java dataset). Any commit with a time later than the release date was output.  This analysis yielded 11 commits from 3 projects in the Java dataset where the dates were in the years 2025, 2027, and 2037.  Clearly these commits have invalid dates. A manual inspection of these commits showed the commits were (based on the Git graph) in between commits with dates that appear valid, indicating the years were off. Most likely these invalid dates were generated through either user error or misconfigured clocks.

The other three datasets all had similarly small number of future commits, with 75 commits in 1 project for the SourceForge dataset, 0 commits for Kotlin, and 4 commits in 3 projects for Python.  In general it appears future dates are relatively rare, regardless of the programming language or forge.

\findings{
    We found both suspiciously old and suspiciously new commit timestamps in all datasets.  Old timestamps (such as 0) are much more common, while suspiciously new timestamps far out in the future are less common.
}

\subsection{Finding Out-of-order Commits}
\label{subsec:order-commits}

Another possible problem with VCS tools allowing users and tools to set the commit date is that the date specified might \textit{seem} valid, but actually be wrong.  This could lead to a graph where a particular node has a commit date that is actually older than its parent node.  Obviously such a case should not make sense.  This might be due to a misconfigured clock on a particular computer\footnote{\url{https://stackoverflow.com/questions/633353}}, specifying the wrong time zone\footnote{\url{https://stackoverflow.com/questions/52507279}}, or any other number of causes.\footnote{\url{https://stackoverflow.com/questions/16259105}}  We call these \textit{out-of-order commits}.

In this section, we investigate how frequently such out-of-order commits occur in Boa's dataset.  To do this, we traverse the revision list of each code repository in order and compare the commit date of a revision to the commit date of the previous revision.  Due to how Boa linearizes the commit graph using a topological sort, this might not technically be a parent (indeed, commit nodes might also have multiple parents due to branching) but it can give us insight into this problem.

In the first attempt at writing this query, we noticed a lot of results where one of the two commits were explicitly marked (in the log) as a merge commit.  We decided to filter those out as merging behavior might induce a lot of false positives.  The relevant part of the Boa query in \figref{fig:boaquery} is lines \ref{ln:order1}--\ref{ln:order2}.

Running this query on the Java dataset gave us 18,685 suspicious commits from 4,275 projects.  For those projects, this represents 0.59\% of their total commits.  For the full dataset, this represents 0.12\% of the commits.  We used the GitHub API to download the JSON metadata for as many of the commits as possible and for any missing commits attempted to obtain JSON metadata from Software Heritage.  This left us with 18,379 commits, which we then verified their commit timestamp against each of their parent commit timestamps. That process indicated a total of 13,611 commits from 3,967 projects had at least one parent that was newer than the commit itself.

Kotlin had a similar number of out-of-order commits, totalling 2,635 commits from 1,754 projects.  Python actually had about three times more than Java (per project) with 12,275 commits from 1,376 projects.  And SourceForge had substantially more, with 65,238 commits from 53,820 projects.

\begin{table}[ht]
  \centering
  \caption{For out-of-order commits, how far in the future (in seconds) is the parent commit compared to the child commit?}
  \label{tab:order-badness}
  \resizebox{\linewidth}{!}{
\begin{tabular}{lrrcr@{\hskip.2em}!{/}@{\hskip.2em}r@{\hskip.2em}!{/}@{\hskip.2em}rr}
  \toprule
\textbf{dataset} & \textbf{mean} & \textbf{std}   & \textbf{min} & \textbf{25\%} & \textbf{50\%} & \textbf{75\%} & \textbf{max}  \\ 
  \midrule
  Java           & 66,336,948.24 & 279,541,949.43 & 1            & 2,726         & 28,418        & 313,125.25    & 1,395,555,801 \\ 
  Kotlin         & 3,176,263.51  & 47,386,198.06  & 0            & 1,208         & 9,688         & 40,054.00     & 1,585,965,306 \\ 
  Python         & 1,680,997.88  & 33,655,968.10  & 0            & 12,357        & 39,944        & 86,400.00     & 1,485,020,157 \\ 
   \bottomrule
\end{tabular}
}
\end{table}

\begin{figure}[ht]
  \centering
  \includegraphics[width=\linewidth]{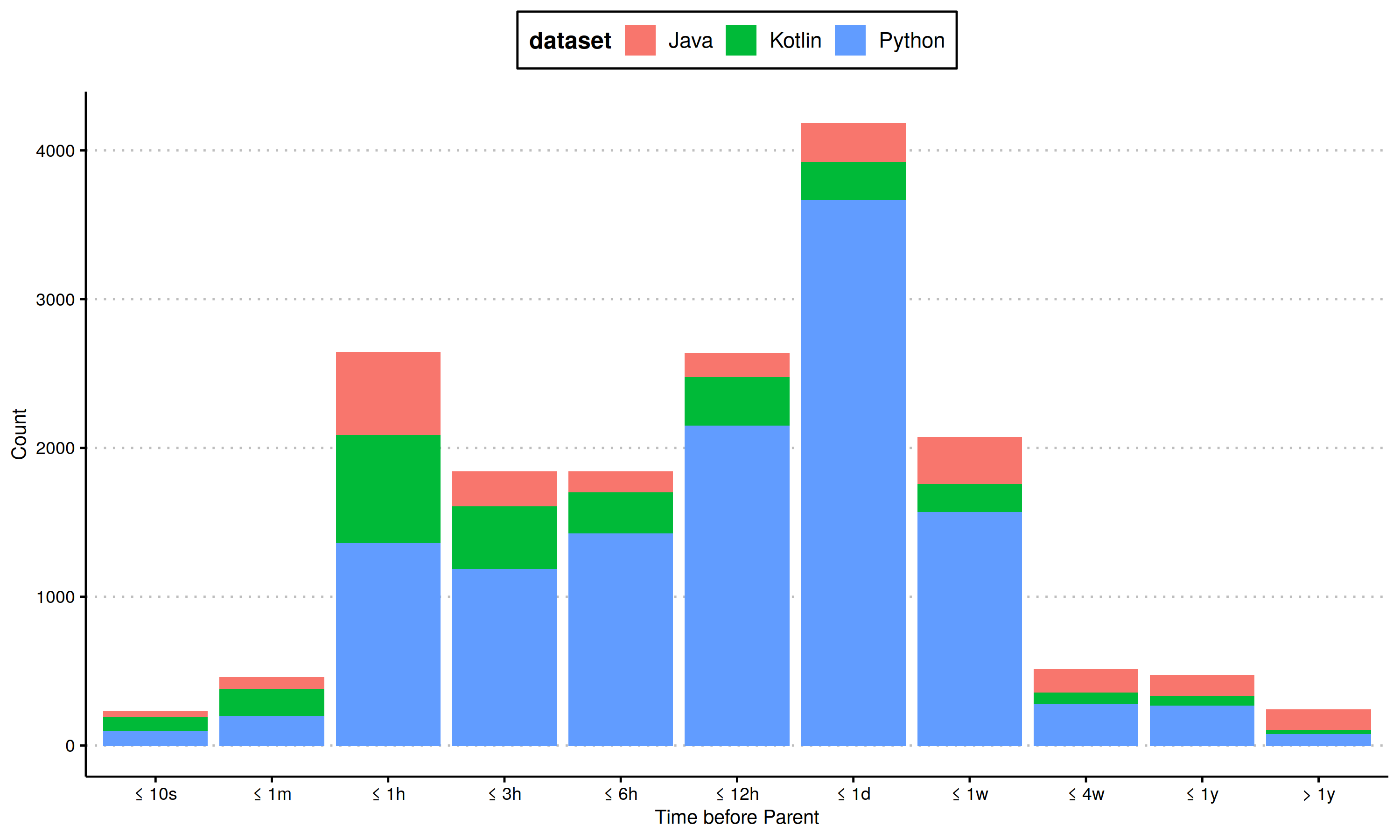}
  \caption{Histogram of \(parent - child\) differences for out-of-order commits.}
  \label{fig:categories-order-difference}
\end{figure}

From these out-of-order commits in all three datasets, we examine just how far the children are committed before their parent.  If the distance is short, it might indicate clock skew issues.  If the distance is far, it might indicate tools causing the problem.  In terms of seconds, we show the summary statistics in \tabref{tab:order-badness}.  This shows that the median out-of-order commit is around 28k seconds before its parent in Java, 9k seconds before its parent in Kotlin, and almost 40k seconds before its parent in Python.  All of these values are less than one day.  In addition to the median (50\%), we present mean, standard deviation, minimum , 25th and 75th percentiles, and maximum.

We also present a histogram of the data as eleven buckets in \figref{fig:categories-order-difference}, from fairly small time spans (\(\leq 30s\)) to medium-size (\(\leq 1d\)) to much longer (\(> 1y\)).  Now you can clearly see that many appear to occur within one day, hinting at a potential misconfigured timezone, and those with a difference less than a minute are likely some level of clock skew.  However this is just speculation, as we can't tell from the commits themselves what accounts for these differences.

\findings{
    All four datasets contained a large number (2k-18k) of out-of-order commits.
}

In the next sections we look at some common tools, users, and projects observed in the out-of-order dataset.

\begin{table*}
\centering
\caption{Twenty most common tokens appearing in bad commit messages.  English stop words were removed.}
\label{tab:wordcloud-order}
\begin{tabular}{r@{\hspace{0.5em}}l@{\hspace{10em}}r@{\hspace{0.5em}}l}
  \toprule
  \textbf{\(n\)} & \textbf{Token} & \textbf{\(n\)} & \textbf{Token} \\
  \midrule
3264 & updat      & 868 & file                                 \\
2744 & fix        & 828 & commit                               \\
1846 & ad         & 800 & reviewed-bi                          \\
1473 & add        & 770 & gener                                \\
1440 & http       & 744 & thi                                  \\
1185 & chang      & 731 & remov                                \\
1148 & git-svn-id & 685 & bug                                  \\
1099 & test       & 637 & code                                 \\
906  & creat      & 574 & d0ab736e-dc22-4aeb-8dc9-08def0aa14fd \\
891  & use        & 573 & work                                 \\
\bottomrule
\end{tabular}
\end{table*}

\subsubsection{Common Tools}

We further processed the commits suspected to be out of order.  Having done so, we collected all commit messages and removed English stop words to produce a table of frequent words (\tabref{tab:wordcloud-order}).  We also generated a word cloud (not shown) of all words, allowing us to visually analyze terms frequently used in the bad commits.  We used the word cloud to get a feel for some commonly occurring problems without limiting ourselves to just the most frequently occurring words.  By doing so, we were able to note a handful of tools that have a tendency to produce bad commit timestamps.

Review systems like Gerrit\footnote{\url{https://gerritcodereview.com}} seem to be a frequent contributor to bad commits, as found by the \texttt{Change-Id} commit footer (511 times).  We suspect this is due to the ``push, review, commit/rebase, force push, GOTO review'' method that is used by many participants in the code review process.

We also frequently found other commit log footers, like \texttt{Reviewed-by} (920 times total).  These are used in other review processes, which involve either rebasing to edit commit messages to include them, or passing patch sets via email.

We also noticed another mixed VCS, namely \texttt{hg-git}\footnote{\url{https://www.mercurial-scm.org/wiki/HgGit}}, which allows a Mercurial user to manipulate a Git repository using Mercurial commands.  In particular, we note the addition of metadata to commits, with the \texttt{rebase\_source} footer (seen 351 times), which is likely a result of rebases on Git repositories using Mercurial or similar tools.  Mercurial's abbreviation, \texttt{hg} is also found 523 times.

Google produced a tool, MOE\footnote{\url{https://opensource.google/projects/moe}} (Make Open Easy) which is used to synchronize two repositories, one internal, and one open to the public.  This tool can synchronize, translate content between kinds of repositories, and scrub content from a repository.  Because of these features, we suspect that use of this tool produced a mismatch between repositories, where an open-source repository received patches from an internal repository after receiving patches from other contributors.  We see MOE related messages show up 465 times across the bad commits.

\findings{
    There are several commonly used tools that seem to cause bad Git timestamps, often tools that convert or interoperate with other version control systems.
}

\subsubsection{Commits On and Off GitHub}

Specifically when we inspected the Kotlin dataset, we noticed a lot of commonly occurring commit messages along the lines of ``Created file.ext'' or ``Updated file.ext''.  We inspected a couple of these results to see what might be happening and quickly realized those repositories had a mixture of verified and unverified commits.

\begin{figure}
    \centering
    \includegraphics[width=\linewidth]{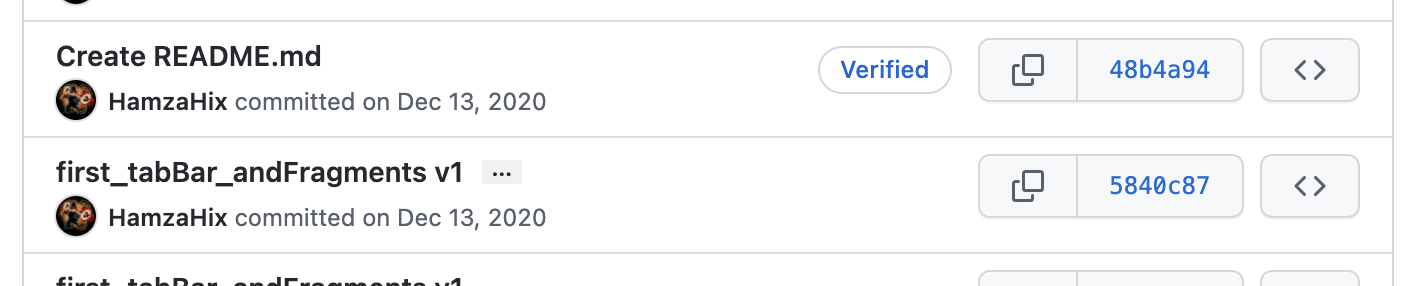}
    \caption{An example of verified and non-verified commits that are out-of-order (from \url{https://github.com/HamzaHix/first_tabBar_andFragments/commits/master}).}
    \label{fig:verified}
\end{figure}

For example, consider the commits shown in \figref{fig:verified}.  The parent commit has a time of 6:39PM CST.  The child commit has an older time of 5:56PM CST.  But notice the child is marked as a verified commit. This commit states ``This commit was created on GitHub.com and signed with GitHub’s verified signature.'' The parent commit does not indicate it was a verified commit, and thus we can safely assume it was created on a different machine and pushed to GitHub.

This seems to indicate the user's machine had the wrong time set: either the time was just bad or there was some timezone misconfiguration causing it to be off by one or more hours.  We investigated how often the out-of-order commits in our dataset were marked as verified by inspecting the commit's JSON data to look for the \texttt{verified} attribute.  In total, we found 864 out-of-order commits were marked as verified.

\findings{
    Machine time differences, between the GitHub servers and individual user's machines, can often introduce bad Git timestamps.
}

\subsubsection{Common Authors and Repositories}

We were interested to see if a few number of authors or repositories contributed a large number of bad commits.  If this is the case, it could make filtering much easier in cases where a small number of bad commits might still be acceptable.

Using the collected commit data, we analyzed commit author information and counted the number of commits made by the top-20 committers of the bad commits in our dataset.  From this, we found that 9,383 commits (26\% of all bad commits) were made by the top 20 committers (with all `(no name)' committers grouped as one).  We keep commits with an unknown committer to help understand possible causes for time issues.  We suspect that a commit made without a committer name is more likely to be using a mis-configured tool.  Since we observed the use of common tools as a common source of bad commit timestamps, it makes sense there may be some committers (who utilize those tools) with a larger number of bad commits.

\begin{table}
\caption{Top 20 projects with the most faulty commits.}
\label{tbl:proj-faulty}
\centering
\begin{tabular}{rl}
\toprule
\textbf{Bad Commits} & \textbf{Project} \\
\midrule
3347 & rawbinz/pythonsnippets \\
3314 & jacksyen/pyastd \\
771 & kylenapped/shashy \\
571 & HansiChan/SoccerPredictor \\
535 & securesystemslab/zippy \\
459 & ghaseminya/commiter \\
195 & joehzli/seattle \\
143 & uditrugman/openjdk8-jdk \\
113 & xapi-project/xen-api \\
100 & mytskine/mupdf-unofficial \\
80 & uditrugman/openjdk8-hotspot \\
69 & D5rkUnl0ck3r/test \\
68 & gisce/openobject-addons-extra \\
60 & cjashfor/LinuxToolsProjectPatches \\
56 & iw3hxn/server \\
51 & cylc/cylc-flow \\
50 & igloosec/hue \\
43 & tandong8888/topsun \\
41 & talknomoney66/TalkNoMoneyShare \\
37 & cbeust/kobalt \\
\bottomrule
\end{tabular}
\end{table}

Similarly, we collected the repository each bad commit belonged to and found that the top-20 repositories, shown in \tabref{tbl:proj-faulty}, contributed 10,103 commits (59\% of all bad commits).  Note that some of these projects appear to be clones (e.g., \texttt{openjdk8-hotspot} and \texttt{openjdk8-jdk}), where the original repository most likely also contains bad commit timestamps.  Boa only contains repositories not explicitly marked as forks (what \citet{pietri20forks} call ``forge forks''), so these projects most likely cloned and uploaded the repository without utilizing GitHub's fork feature.

\findings{
    These results indicate a small number (20) of authors and repositories tend to account for a relatively large percentage (26\% and 59\%, respectively) of bad commits.
}

\subsection{Summary}

To summarize, we were able to find thousands of bad commit timestamps, as shown in \tabref{tab:bad-timestamps-summary}.  All four datasets contained bad timestamps, with the older SourceForge data having more than the newer Git datasets.  All programming languages seem to contain commits with bad time data.

\begin{table}[htbp]
  \centering
  \caption{Summary of commits exhibiting ``bad'' timestamps.}
  \label{tab:bad-timestamps-summary}
  \resizebox{0.95\linewidth}{!}{\begin{tabular}{lrrrrrr}
  \toprule
              & \multicolumn{2}{c}{\textbf{Old}} & \multicolumn{2}{c}{\textbf{Out-of-order}} & \multicolumn{2}{c}{\textbf{Future}}                                         \\
              & \textbf{Commits}                 & \textbf{Projects}                         & \textbf{Commits} & \textbf{Projects} & \textbf{Commits} & \textbf{Projects} \\
  \midrule
  SourceForge & 46,266                           & 11                                        & 65,238           & 53,820            & 75               & 1                 \\
  Java        & 3,612                            & 51                                        & 13,599           & 3,967             & 11               & 3                 \\
  Kotlin      & 6                                & 6                                         & 2,635            & 1,754             & 0                & 0                 \\
  Python      & 20                               & 8                                         & 12,275           & 1,376             & 4                & 3                 \\ 
  \bottomrule
\end{tabular}
}
\end{table}

Many of these commits seem to originate from tool use, especially tools that migrate or synchronize between two version control systems.  In addition, we saw a small number of committers and projects seem to contribute a large number of the bad commit timestamps.

\section{Potential Impact of Time Problems}
\label{sec:impact}

To investigate the potential impact the time issues we observed might have, we chose to look at the published MSR data papers in detail to see if there may be time data problems with those datasets.  We did not investigate the data from the technical track papers, as many research papers (especially older ones) lack making the data available or the link to their data is no longer valid~\citep{robles10}.  Additionally, the data papers have potential for lots of reuse, as was shown by \citet{kotti19}.  Thus if there are problems with the data in the data papers, potentially many other papers might be affected.

\subsection{Additional Filtering}

First we looked through the 81 data papers and identified any additional filtering (beyond time-based filtering) they might have performed, as such a filter might potentially weed out bad time-based data as well.  Two of the authors independently read each paper's approach and evaluation sections to look for any mention of potential filtering, such as filtering projects based on metadata or commits based on deduplication.  They then discussed any differing results to reach an agreement.  Based on that analysis, it seems many data papers (48 out of 81) do not explicitly mention any filtering.  The top three filtering techniques observed were: deduplication (at various levels) in 7 (8.64\%) papers, popularity (stargazer count on GitHub) in 5 (6.17\%) papers, and selecting specific programming languages (often, Java) in 3 (3.70\%) papers.

The datasets we analyzed in this paper were all deduplicated and spanned several programming languages, including Java, Kotlin, and Python.  Despite that, we still observed quite a few bad commits, so it seems these two filtering techniques are probably not sufficient to avoid bad time-based data.  Given that result, we focus here on using stargazer counts to filter bad time-based data.

Based on a range of noted minimum star counts from the selected dataset papers, we report the number of commits remaining when we filter projects by a minimum of stars across the three GitHub-derived Boa datasets.  The results are shown in \tabref{tbl:filtering-stars}.

\begin{table}[h]
  \centering
  \caption{Bad commits remaining in projects when filtered by stars.}
  \label{tbl:filtering-stars}
  \begin{tabular}{rr@{\hskip.2em}!{/}@{\hskip.2em}lr@{\hskip.2em}!{/}@{\hskip.2em}lr@{\hskip.2em}!{/}@{\hskip.2em}l}
    \toprule
    \textbf{Min. Stars} & \multicolumn{2}{c}{\textbf{Java}} & \multicolumn{2}{c}{\textbf{Kotlin}} & \multicolumn{2}{c}{\textbf{Python}} \\
    \midrule
    1                   & 11,460                            & 51\%                                & 967 & 37\% & 3,740 & 30\%           \\
    2                   & 10,339                            & 46\%                                & 672 & 25\% & 2,949 & 24\%           \\
    5                   & 8,663                             & 39\%                                & 474 & 18\% & 2,001 & 16\%           \\
    10                  & 5,844                             & 26\%                                & 367 & 14\% & 1,787 & 15\%           \\
    50                  & 2,805                             & 13\%                                & 219 & 8\%  & 1,538 & 13\%           \\
    \textbf{100}        & \textbf{2,365}                    & \textbf{11\%}                       & \textbf{165} & \textbf{6\%}  & \textbf{895}   & \textbf{7\%} \\
    500                 & 385                               & 2\%                                 & 52  & 2\%  & 120   & 1\%            \\
    1000                & 216                               & 1\%                                 & 36  & 1\%  & 86    & 0.7\%          \\
    \bottomrule
\end{tabular}

\end{table}

These results show that even with the highest stargazer filter we found in the data papers (100, from \citeappL{9463133}), thousands of commits with time-based problems still remain.  In general, it seems possible to filter a large percent of bad commits using stars, but would require a higher cutoff value than typically observed (such as 500).  We have personally observed some research papers that select the top-K projects based on highest star counts.

Such an approach may be effective, for example in Java the top-1k projects by star count would yield projects with over 2k stars each and, in theory, filter a large amount of bad time-based data.  This however is no guarantee, as one of the projects in our datasets that contained bad time data (the Python project \texttt{scrapy}\footnote{\url{https://github.com/scrapy/scrapy}}, with 17 bad commits) actually has over 41k stars.

\findings{
    Filtering on non-time based criteria, such as stargazer count, often is not effective.  It can be effective, if the cutoff is sufficiently high (500+) or when selecting the top-K projects sorted by star count.
}

\subsection{Investigating Data Papers}

Since it seems likely the data papers are not employing a filtering technique that would filter out bad time data and since we have several Git-based datasets, here we look at all data papers that used Git data.  This gave us 34 papers out of the 81 published data papers using time-based data.  From that set of 34, we then checked if the dataset is still accessible and if it provides raw Git repositories.  This left us with 11 out of 81 data papers (13.58\%) for our analysis.

\setlength{\rotFPtop}{0pt plus 1fil}
\begin{sidewaystable}
  \centering
  \caption{MSR data papers analyzed for potentially bad time data.}
  \label{tab:datapapers-problems}
  \begin{tabular}{lrlr@{\hskip.2em}!{/}@{\hskip.2em}l}
  \toprule
  \textbf{Dataset Paper}                                    & \textbf{Citations} & \textbf{Boa Dataset(s)} & \multicolumn{2}{c}{\textbf{Projects Covered}} \\
                                                            &                    &                         & \textbf{In Boa}   & \textbf{Original Dataset} \\ 
  \cmidrule(r){1-1}\cmidrule(l){2-5}
  Diversity~\citepappF{10.5555/2820518.2820601}             & 79                 & Java, Kotlin, Python    & 1,539             & 23,474                    \\
  AndroidTimeMachine~\citepappI{8595172}                    & 37                 & Java, Kotlin            & 1,087             & 8,216                     \\
  Docker~\citepappI{8595171}                                & 20                 & Java, Kotlin, Python    & 353               & 98,033                    \\
  DoSC~\citepappH{7962412}                                  & 14                 & Java, Kotlin, Python    & 6                 & 10                        \\
  AndroZoo~\citepappK{10.1145/3379597.3387503}              & 6                  & Java, Kotlin            & 5,565             & 50,217                    \\
  Enterprise-driven OSS~\citepappK{10.1145/3379597.3387495} & 4                  & Java, Kotlin, Python    & 487               & 17,255                    \\
  SEART~\citepappL{9463094}                                 & 2                  & Java, Kotlin, Python    & 13,705            & 938,510                   \\
  DUETS~\citepappL{9463096}                                 & 1                  & Java                    & 13,172            & 147,991                   \\
  JTeC~\citepappK{10.1145/3379597.3387484}                  & 1                  & Java                    & 12,906            & 31,232                    \\
  Wonderless~\citepappL{9463099}                            & 0                  & Java, Kotlin, Python    & 3                 & 1,877                     \\
  GE526~\citepappL{9463127}                                 & 0                  & Java, Kotlin, Python    & 2                 & 101                       \\
  \bottomrule           
\end{tabular}

  \vspace{2em}
  \centering
  \caption{Summary of commits exhibiting ``bad'' timestamps in studied MSR data papers.}
  \label{tab:datapapers-problems2}
  \begin{tabular}{lrrrrrr}
  \toprule
                                                            & \multicolumn{2}{c}{\textbf{Old}}     & \multicolumn{2}{c}{\textbf{Out-of-order}} & \multicolumn{2}{c}{\textbf{Future}}  \\
  \textbf{Dataset Paper}                                    & \textbf{Commits} & \textbf{Projects} & \textbf{Commits} & \textbf{Projects}      & \textbf{Commits} & \textbf{Projects} \\
  \cmidrule(r){1-1}\cmidrule(l){2-7}
  Diversity~\citepappF{10.5555/2820518.2820601}             & 0                & 0                 & 921              & 114                    & 0                & 0                 \\ 
  AndroidTimeMachine~\citepappI{8595172}                    & 0                & 0                 & 723              & 30                     & 0                & 0                 \\
  Docker~\citepappI{8595171}                                & 0                & 0                 & 99               & 21                     & 0                & 0                 \\ 
  DoSC~\citepappH{7962412}                                  & 0                & 0                 & 4                & 1                      & 0                & 0                 \\ 
  AndroZoo~\citepappK{10.1145/3379597.3387503}              & 1                & 1                 & 265              & 94                     & 0                & 0                 \\ 
  Enterprise-driven OSS~\citepappK{10.1145/3379597.3387495} & 2                & 1                 & 337              & 41                     & 0                & 0                 \\ 
  SEART~\citepappL{9463094}                                 & 985              & 12                & 3,842            & 500                    & 0                & 0                 \\
  DUETS~\citepappL{9463096}                                 & 127              & 8                 & 2,005            & 424                    & 0                & 0                 \\ 
  JTeC~\citepappK{10.1145/3379597.3387484}                  & 2,896            & 14                & 2,593            & 553                    & 0                & 0                 \\ 
  Wonderless~\citepappL{9463099}                            & 0                & 0                 & 0                & 0                      & 0                & 0                 \\
  GE526~\citepappL{9463127}                                 & 0                & 0                 & 0                & 0                      & 0                & 0                 \\
  \bottomrule
\end{tabular}

\end{sidewaystable}

For each of the 11 data papers, we intersected the projects in their dataset against the Java, Kotlin, and Python datasets to see if we have similar projects.  The results are shown in \tabref{tab:datapapers-problems}, where we list the paper, how many citations it has at the time of writing (based on Google Scholar), which Boa dataset(s) intersected with it, and the ``Projects Covered'' column contains two numbers: the first is the number of projects intersecting the Boa dataset(s) and the second is the total number of GitHub projects in the original data paper.

Then, based on the list of projects we found, we look to see how many commits in those projects contained time errors.  The results are shown in \tabref{tab:datapapers-problems2}.  As in the previous sections, these results are all deduplicated commits.

First, we notice that in 9 of the 11 datasets we were able to find bad commits. This does not mean the other 2 datasets lack bad time data, simply that the intersection was very small and contained no bad commits and thus we are unable to easily identify problems without inspecting each project in those datasets.  All of the other 9 projects contained out-of-order commits and none of them contained future commits.  5 of the 9 contained old commits. The fact we found so many problems with such a small sample of the datasets (anywhere from 0.36\% to 40.04\% of the projects) hints that there could be more bad commits in those datasets.

These 9 papers have been cited 164 times according to Google Scholar at the time of writing this paper.  Thus there are a large number of research papers that potentially used these datasets and may have relied on these bad commit timestamps. This shows that it is especially important to carefully handle (filter and/or clean) commit data when building a reusable research dataset as any potential problems with the data could propagate to other research.

\findings{
    When looking at 0.36\% to 40.04\% of the projects contained in 11 previously published dataset papers, we found 9 of the 11 papers that are cited 164 times contain bad time data.
}

\section{Discussion and Guidelines}
\label{sec:discussion}

We showed that time-based data is utilized by a large number of MSR research (at least 38\% of papers).  We then described some possible problems with timestamps in Git data, the most used data kind, and attempted to quantify how often those problems occur in the most used data source, GitHub.  In this section, we discuss some guidelines for handling time-based data.

When order is a component of an analysis, handling suspicious commits is recommended.  To do this, we recommend that any commit with a timestamp less than 1 is removed.  For the data we observed, this filter would remove about 98\% of suspicious commits.

In general, we also recommend searching the commit logs for projects that contain a `\texttt{git-svn-id}' tag and consider removing matching projects.

To handle the problem of out-of-order commits, we recommend four strategies: \begin{enumerate*}
\item filtering commits before a certain date;
\item filtering commits belonging to certain projects;
\item filtering only commits which are out-of-order; and/or,
\item using a robust method of mining commits with rebasing.
\end{enumerate*} We will discuss each of these in turn, including the benefits and the potential problems each brings to the table.

\subsection{Filtering Before a Specific Date}

\begin{table}
\caption{Percent of faulty commits removed by filtering commits from or before a given year.}
\label{tbl:filtering-by-year}
\centering
\begin{tabular}{crr}
\toprule
\textbf{Filter Date} & \textbf{Removed} & \textbf{Bad Commits Remaining} \\
\midrule
$\leq 2015$ & 99.45\% &    11 \\
$\leq 2014$ & 99.29\% &    14 \\
$\leq 2013$ & 97.73\% &    45 \\
$\leq 2012$ & 70.38\% &   593 \\
$\leq 2011$ & 50.69\% &   987 \\
$\leq 2010$ & 36.17\% & 1,278 \\
$\leq 2009$ & 24.86\% & 1,504 \\
$\leq 2008$ & 18.14\% & 1,639 \\
$\leq 2007$ & 12.84\% & 1,745 \\
$\leq 2006$ & 11.05\% & 1,781 \\
$\leq 2005$ & 10.03\% & 1,801 \\
$\leq 2004$ &  8.95\% & 1,823 \\
$\leq 2003$ &  8.15\% & 1,839 \\
\multicolumn{3}{c}{...}\\
$\leq 1992$ &  6.63\% & 1,869 \\
\multicolumn{3}{c}{...}\\
$\leq 1970$ &  6.63\% & 1,869 \\
\bottomrule
\end{tabular}

\end{table}

The first filtering method we suggest is removing commits from \emph{before} a specific date.  We suggest this method due to its relative simplicity, as well as its effect (see \tabref{tbl:filtering-by-year}).  In particular, we suggest removing all commits before 1 January 2014, as doing so could remove 97.73\% of all bad commits.  Even filtering this much data still leaves (at the time of writing) eight years of historical data to study.

If the research question requires longer history, this method may not be feasible and the other filtering methods mentioned later are recommended.

\subsection{Filtering Specific Projects}

Given the exceedingly wide range of projects available on GitHub and similar sites, the removal of projects known to have a large number of out-of-order commits still leaves a large available corpus.  The benefit is that a longer history can be maintained.  This is a bit more complicated than simply filtering by a specific date though, as a list of so called ``bad'' projects would need to be known.  The MSR community could work toward maintaining such a list.

\subsection{Filtering only Out-of-Order Commits}

Another recommendation is to filter the out-of-order commits.  The specific analysis would have to decide which commit(s) to remove in this case, and how that might affect the analysis results.  For example, the simplest solution would be to remove all out-of-order commits.  However if an analysis relies on pairs of commits, e.g. to determine co-changes, such a study may have to remove additional neighboring commits.  Similarly, if a study looks at the full history of files, then any files modified in the removed commits may trigger additional removals of other commits modifying the same set of files.  In such a case, it may be easier to simply reject any project with one or more out-of-order commit.

This filtering method is the most computationally expensive, requiring each commit to be examined in turn. However, this method has the benefit of not removing other history and could be used to enable study of a repository that otherwise may be problematic, or has substantial pre-2014 history.  While some previous work, such as \citet{steff12}, built graphs with commit timestamps and performed topological sorts on them and could very easily have identified such out-of-order commits, they did not identify such a problem or suggest it as a solution.

\subsection{Ordering on Committer Date}

Git commits store two timestamps: the \textbf{author date} and the \textbf{committer date}.  The author date is the date the commit was originally made.  The committer date is essentially the last date the commit was modified.  Certain commands, e.g. cherry picking, amending, and rebasing, modify the commits and thus will change the committer date.  Most of the time, the two timestamps are identical.

In this work we looked for out-of-order commits using the committer date.  As Git is a graph, those committer timestamps should be in non-decreasing order as you walk the graph.  The author timestamps however might be out of order -- and that is to be expected.

We recommend most papers utilize the committer date, and then look for out-of-order issues in a project.  For any work that needs to know about rebasing, e.g. when analyzing the code review process, they may have to apply more complicated techniques, especially if the project relies on a rebase-heavy workflow, such as those utilizing Gerrit or similar review platforms.

\subsection{Using Topological Order}

Another possibility to handle out-of-order commits would be to ignore the timestamps entirely and rely only on the actual structure of the graph.  A topological ordering of that graph would provide the correct order of the commits, regardless of their individual timestamps.  Depending on the particular analysis used, this strategy might be sufficient.  This would only work if the analysis does not rely on the specific timestamps and only needs to know the relationship among the commits (parent-child, etc).  For example, any analysis that simply needs to know about the diff between two versions of source files (e.g., if mining for refactorings).

\subsection{Filtering by Star Counts}

As we observed in \secref{sec:impact}, star counts are not always a good method for filtering out bad time data.  It can be tricky to pick a sufficiently high cutoff value, without selecting a high value of 1000 or more.

One method that probably works well however is to simply select the top-K projects, based on star counts.  Often, as long as the researcher is only selecting the top 1k or so projects, those projects will have a high enough star count to filter out much of the bad data.

\subsection{Summary}

The exact method(s) utilized will depend highly on the specific research questions being answered.  For example, if the research questions require a long history then filtering by date might not be the best approach.  Additionally, researchers need to decide if it is acceptable to simply drop the bad commits, or if projects with even a single bad commit should be excluded entirely.

\section{Threats to Validity}
\label{sec:threats}

A threat to construct validity is the use of keyword-based search technique to identify MSR papers using time-based data. We find this technique is sound, as we manually verified the results.  It is however not complete, as a paper might have utilized time-based data without using any of the selected keywords.  Thus the 38\% of papers identified is a lower bound and the actual number of papers utilizing time-based data might be even higher. The percentage we found is still high enough to indicate this is an important problem.

Another threat to construct validity is the choice to exclude merge commits when looking for out-of-order commits.  This was done as we observed quite a few false-positives.  However, many merge commits might actually contain out-of-order timestamps and thus our choice might lead to under counting the amount of bad time-based commits.  We chose to err on the side of precision at the expense of recall.

The notes for the ``2019 October/GitHub'' dataset indicate that the data was actually collected in 2015\footnote{\url{http://boa.cs.iastate.edu/boa/?q=content/dataset-notes-october-2019}}.  Because of the age of the data, many projects no longer exist on GitHub. Even with the newer Kotlin and Python datasets there are several projects no longer available.  To mitigate this threat, we utilized the Software Heritage Archive~\citep{software-heritage,software-heritage-archive} to attempt to locate these repositories which have been deleted.

Boa's datasets exclude explicitly marked forks (what \citetappK{msr20-27} call a ``forge fork'').  However, forks created off the website remain in the dataset causing some commits to be duplicated.  We attempted to mitigate this threat by identifying and removing exact duplicate commits (same commit hash) from the GitHub datasets.

A threat to internal validity is that some of the commits might actually suffer from multiple problems.  To quantify how many commits have a bad timestamp and also are out of order, we intersect the two results.  There are 877 (4.09\%) commits in Boa's dataset that potentially suffered from both the out-of-order error as well as being suspiciously old.  Most of those (all but 11) are timestamps of 0.

For Python we actually see slightly different results, as only 6 (0.042\%) commits in Boa's dataset potentially suffered from both the out-of-order error as well as being suspiciously old.  Out of those 6, only 1 is the timestamp 0.

For Kotlin the results are similar to Python, as only 4 (0.14\%) commits in Boa's dataset potentially suffered from both the out-of-order error as well as being suspiciously old.  None of those commits are the timestamp 0.  Overall, it appears that very few commits in the datasets we studied suffer from multiple problems.

A threat to internal validity is that Boa's datasets tend to include a lot of duplicated data (exact code clones). This often occurs due to a fork occurring outside of GitHub that is not a ``forge fork'' (using the terminology of \citetappK{msr20-27}).  We attempted to mitigate this issue by removing exactly duplicated commits (commits with the same hash) and then reporting results for only the deduplicated data. We were only able to do this for the three GitHub-based datasets, so the SourceForge dataset might have duplicated commits in it.

Another threat to internal validity is that 917 out of the 7,569 projects (12.12\%) identified with time data problems no longer exist on GitHub (as of 8 October 2021).  These projects are however still in the released Boa dataset, so we maintained their results in this study.  We attempted to mitigate this threat by validating the data directly with GitHub using their API, and for the projects that were missing we utilized the Software Heritage Archive (all projects were found).  Note however that the times Boa and Software Heritage indexed the projects might differ, and thus there were some commits found by Boa that we were not able to verify. Such commits were removed from the dataset.  Of the 38,624 total Git commits found by Boa, 33,898 (87.76\%) were still on GitHub, 3,572 (9.25\%) were found on Software Heritage, and 1,154 (2.99\%) were excluded.  We do not believe this is a problem however, as the point of the analysis was to see if bad time data exists, not to fully account for all such cases.

A threat to external validity is that we only originally studied Git repositories. However, although we did not quantify it explicitly, Subversion also allows developers to modify (and even remove!) the commit date: \texttt{svn propset -rXXX --revprop svn:date}.  This was why we also analyzed an older SourceForge dataset that contains CVS and SVN repositories and saw similar issues as observed in the GitHub datasets.  The results however may not generalize to any VCS that disallows modifying commit timestamps.

Another threat to external validity is that certain tools that cause time problems are only available for certain programming languages so there is a need to study commit data from multiple languages.  To mitigate this threat, we studied three datasets from GitHub from three popular languages: Java, Kotlin, and Python.  We also studied SourceForge, which contains commit data from over 50 programming languages, including over 23k Java projects, 9k C++ projects, 4k C projects, 4k C\# projects, and 900 PHP projects.

\section{Conclusion}
\label{sec:conclusion}

The use of time-based data in MSR studies is wide-spread, in at least 38\% of MSR papers.  Properly handling this time-based data is thus very important.  However, the diversity of tools and workflows used to generate the time-based data can present challenges.  In particular, ensuring that time data is consistent and maintains linearity is important.  Further, we have found that many papers do not describe cleaning or filtering of time data, with those that do describe filtering tending towards simple techniques like selection from a defined time span or selection of data before a certain date.  Some papers have used more robust or rigorous techniques, and may thus avoid some of the time-related problems found in data such as Git repositories.

To remedy potential time-based issues in VCS data, such as that coming from GitHub, we recommend a simple filter to drop any timestamp less than 1 as well as a more complex filtering to remove out-of-order commits.  Ideally, each repository would be analyzed to detect and remove the out-of-order commits, but depending on the need a simple cutoff filter removing commits prior to 2014 might suffice.  Applying both filters (the second filter actually implies the first) is very simple and would remove around 98\% of all observed bad commits.

We also observed 9 (out of 11 studied) MSR data papers that exhibit bad time data, showing that the problems identified occur in practice. These papers already have over 150 citations, indicating further research has utilized this potentially bad data.

In the future, we would like to investigate potential problems in other kinds of time data, such as issue reports.  We would also like to investigate how time-based data is used when training machine learning models and if issues arise from training on later observed data and then classifying on older data.

\begin{acknowledgements}
The authors would like to thank Yijia Huang, Tien N. Nguyen, and Hridesh Rajan for insightful discussions that inspired this paper.  We also thank the anonymous MSR'21 and EMSE reviewers for many suggestions that substantially improved this paper.
\end{acknowledgements}

\section*{Conflict of interest}
The authors declare that they have no conflict of interest.

\bibliographystyle{spbasic}
\bibliography{refs}

\clearpage
\appendix

\section{List of Boa jobs}
\label{app:boa}

In this section we list public links to all of the Boa queries utilized by our study.  Full details as well as all data (including generated data based on the Boa outputs) is available in our replication package~\citep{replication}.

\subsection{Java Queries}
All Boa queries were run on the `2019 October/GitHub` dataset.

\vspace{1em}
\noindent Suspiciously 'old' commits: \url{http://boa.cs.iastate.edu/boa/?q=boa/job/public/90164}

\noindent Suspiciously 'future' commits: \url{http://boa.cs.iastate.edu/boa/?q=boa/job/public/90973}

\noindent Out-of-order commits: \url{http://boa.cs.iastate.edu/boa/?q=boa/job/public/90169}

\subsection{Kotlin Queries}
All Boa queries were run on the `2021 Aug/Kotlin` dataset.

\vspace{1em}
\noindent Suspiciously 'old' commits: \url{http://boa.cs.iastate.edu/boa/?q=boa/job/public/95104}

\noindent Suspiciously 'future' commits: \url{http://boa.cs.iastate.edu/boa/?q=boa/job/public/95113}

\noindent Out-of-order commits: \url{http://boa.cs.iastate.edu/boa/?q=boa/job/public/95107}

\subsection{Python Queries}
All Boa queries were run on the `2021 Aug/Python` dataset.

\vspace{1em}
\noindent Suspiciously 'old' commits: \url{http://boa.cs.iastate.edu/boa/?q=boa/job/public/95105}

\noindent Suspiciously 'future' commits: \url{http://boa.cs.iastate.edu/boa/?q=boa/job/public/95112}

\noindent Out-of-order commits: \url{http://boa.cs.iastate.edu/boa/?q=boa/job/public/95106}

\subsection{SourceForge Queries}
All Boa queries were run on the `2013 September/SF` dataset.

\vspace{1em}
\noindent Suspiciously 'old' commits: \url{http://boa.cs.iastate.edu/boa/?q=boa/job/public/95171}

\noindent Suspiciously 'future' commits: \url{http://boa.cs.iastate.edu/boa/?q=boa/job/public/95173}

\noindent Out-of-order commits: \url{http://boa.cs.iastate.edu/boa/?q=boa/job/public/95170}

\section{List of Selected Papers}
\label{app:papers}

In this section, we list all papers selected for inclusion in the study (see \tabref{tab:years}).

\nociteappA{*}
\nociteappB{*}
\nociteappC{*}
\nociteappD{*}
\nociteappE{*}
\nociteappF{*}
\nociteappG{*}
\nociteappH{*}
\nociteappI{*}
\nociteappJ{*}
\nociteappK{*}
\nociteappL{*}
\nociteappM{*}
\nociteappN{*}
\nociteappO{*}
\nociteappP{*}
\nociteappQ{*}
\nociteappR{*}

\bibliographystyleappM{spbasic}
\bibliographyappM{msr2004}

\bibliographystyleappN{spbasic}
\bibliographyappN{msr2005}

\bibliographystyleappO{spbasic}
\bibliographyappO{msr2006}

\bibliographystyleappP{spbasic}
\bibliographyappP{msr2007}

\bibliographystyleappQ{spbasic}
\bibliographyappQ{msr2008}

\bibliographystyleappR{spbasic}
\bibliographyappR{msr2009}

\bibliographystyleappA{spbasic}
\bibliographyappA{msr2010}

\bibliographystyleappB{spbasic}
\bibliographyappB{msr2011}

\bibliographystyleappC{spbasic}
\bibliographyappC{msr2012}

\bibliographystyleappD{spbasic}
\bibliographyappD{msr2013}

\bibliographystyleappE{spbasic}
\bibliographyappE{msr2014}

\bibliographystyleappF{spbasic}
\bibliographyappF{msr2015}

\bibliographystyleappG{spbasic}
\bibliographyappG{msr2016}

\bibliographystyleappH{spbasic}
\bibliographyappH{msr2017}

\bibliographystyleappI{spbasic}
\bibliographyappI{msr2018}

\bibliographystyleappJ{spbasic}
\bibliographyappJ{msr2019}

\bibliographystyleappK{spbasic}
\bibliographyappK{msr2020}

\bibliographystyleappL{spbasic}
\bibliographyappL{msr2021}

\end{document}